\documentclass[11pt,a4paper]{article}

\usepackage{amssymb}
\usepackage{amsmath} 
\usepackage{oldgerm}
\usepackage{amsfonts}
\usepackage{euscript}
\usepackage[dvips]{epsfig}
\usepackage{here}

\begin{document}

\newcommand{\vr}{{\mathbf r}}
\newcommand{\vrt}{\tilde{\mathbf{r}}}
\newcommand{\vrp}{{\mathbf r'}}
\newcommand{\vy}{{\mathbf y }}
\newcommand{\vyp}{{\mathbf y}'}
\newcommand{\be}{\begin{equation}}
\newcommand{\en}{\end{equation}}
\newcommand{\ew}{\epsilon_{\rm w}}
\newcommand{\esolv}{\epsilon_{\rm s}}
\newcommand{\vw}{v_{\rm w}}
\newcommand{\hvw}{\hat{v}_{\rm w}}
\newcommand{\vb}{v_{\rm {\scriptscriptstyle B}}}
\newcommand{\vvw}{v_{\rm w} (\vr;\vrp)}
\newcommand{\dw}{\Delta_{\rm el}}
\newcommand{\rw}{\rho_{\rm w} (x)}
\newcommand{\kd}{\kappa_{\scriptscriptstyle D}}
\newcommand{\zb}{\overline{z}}
\newcommand{\zbu}{\overline{z}^{[1]}}
\newcommand{\zbd}{\overline{z}^{[2]}}
\newcommand{\zbj}{\overline{z}^{[{\rm j}]}}
\newcommand{\zbau}{\overline{z}_\alpha^{[1]}}
\newcommand{\zbad}{\overline{z}_\alpha^{[2]}}
\newcommand{\zbgu}{\overline{z}_\gamma^{[1]}}
\newcommand{\zbgd}{\overline{z}_\gamma^{[2]}}
\newcommand{\zbaj}{\overline{z}_\alpha^{[j]}}
\newcommand{\zbabu}{\overline{z}_{\alpha{\rm b}}^{[1]}}
\newcommand{\zbabd}{\overline{z}_{\alpha{\rm b}}^{[2]}}
\newcommand{\roaw}{\rho_{\alpha }(x;\beta,\{z_\gamma \})}
\newcommand{\roawn}{\rho_{\alpha  {\rm w}}}
\newcommand{\cp}{{\cal P}}
\newcommand{\cg}{{\mathbb{G}}}
\newcommand{\cpp}{{\mathbb{P}}}
\newcommand{\cs}{{\mathbb{S}}}
\newcommand{\fu}{\phi_1}
\newcommand{\fd}{\phi_2}
\newcommand{\usr}{U_{\scriptscriptstyle{\rm SR}}}
\newcommand{\Uelec}{U_{\rm elect}}
\newcommand{\vsr}{v_{\scriptscriptstyle{\rm SR}}}
\newcommand{\Vsr}{V_{\scriptscriptstyle{\rm SR}}}
\newcommand{\hVsr}{\hat{V}_{\scriptscriptstyle{\rm SR}}}
\newcommand{\Vself}{V_{\scriptscriptstyle{\rm self}}}
\newcommand{\hVself}{\hat{V}_{\scriptscriptstyle{\rm self}}}
\newcommand{\Vselfsc}{\Vself^{{\sssc}}}
\newcommand{\Vselfsce}{\Vself^{{\sssc} \, *}}
\newcommand{\row}{\rho_{\alpha \, {\rm w}}}
\newcommand{\roa}{\rho_{\alpha}}
\newcommand{\roab}{\rho_{\alpha}^{ {\scriptscriptstyle\rm B}}}
\newcommand{\fcc}{f^{\rm c \, c}}
\newcommand{\Fcc}{F^{\rm c \, c}}
\newcommand{\Fcm}{F^{\rm c \, m}}
\newcommand{\Fmc}{F^{\rm m \, c}}
\newcommand{\Fmm}{F^{\rm m \, m}}
\newcommand{\ft}{f_{\scriptscriptstyle{\rm T}}}
\newcommand{\Frt}{F_{\scriptscriptstyle{\rm RT}}}
\newcommand{\Ir}{I_{\rm r}}
\newcommand{\rr}{{\EuScript{R}}}
\newcommand{\kud}{\kappa_1^2}
\newcommand{\kdd}{\kappa_2^2}
\newcommand{\kudb}{{\overline{\kappa_1}}^2}
\newcommand{\kddb}{{\overline{\kappa_2}}^2}
\newcommand{\ead}{e_\alpha^2}
\newcommand{\xt}{\tilde{x}}
\newcommand{\vq}{{\bf q}}
\newcommand{\kz}{\kappa_{\rm z}}
\newcommand{\fj}{\phi_{{\rm j}}}
\newcommand{\phit}{\widetilde{\phi}}
\newcommand{\fjt}{\tilde{\phi}_{\rm j}}
\newcommand{\fut}{\tilde{\phi}_{1}}
\newcommand{\fjtp}{\tilde{\phi}_{\rm j}^{*}}
\newcommand{\kzj}{\kappa_{j}}
\newcommand{\kzu}{\kappa_{1}}
\newcommand{\kzd}{\kappa_{2}}
\newcommand{\fub}{\phi_{1{\scriptscriptstyle \rm B}}}
\newcommand{\xxq}{(\xt,\xt', \vq)}
\newcommand{\eq}{E_{\rm q}}
\newcommand{\uj}{U_{\rm j}}
\newcommand{\Uj}{U_{\rm j}}
\newcommand{\hjp}{h_{{\rm j} }^{+*}}
\newcommand{\hjps}{h_{{\rm j} }^{+}}
\newcommand{\hjm}{h_{{\rm j} }^{-*}}
\newcommand{\hjms}{h_{{\rm j} }^{-}}
\newcommand{\bjt}{\tilde{b}_{{\rm j}}}
\newcommand{\bt}{\tilde{b}}
\newcommand{\rac}{\sqrt{1+q^2}}
\newcommand{\racm}{\sqrt{t^2-1}}
\newcommand{\Hjp}{H_{{\rm j}}^{+}}
\newcommand{\Hjm}{H_{{\rm j}}^{ -}}
\newcommand{\Hjme}{H_{{\rm j}}^{ - *}}
\newcommand{\eps}{\varepsilon}
\newcommand{\lt}{\tilde{l}}
\newcommand{\Ee}{\underset{\eps \rightarrow 0}{\rm Exp}}
\newcommand{\El}{\underset{l \rightarrow 0}{\rm Exp}}
\newcommand{\Elt}{\underset{\lt \rightarrow 0}{\rm Exp}}
\newcommand{\Eltseinf}{\underset{ \lt/\eps \rightarrow  \infty}{\rm Exp}}
\newcommand{\Eltinf}{\underset{ \lt \rightarrow  \infty}{\rm Exp}}
\newcommand{\Eu}{\underset{u \rightarrow 0}{\rm Exp}}
\newcommand{\Ujl}{U_{{\rm j}}^{{\rm lin}}}
\newcommand{\Ujlin}{U_{{\rm j}}^{{\rm lin}}}
\newcommand{\Od}{{\cal O}_{{\rm exp}} (\eps^2)}
\newcommand{\fz}{\phi(\dw=0)}
\newcommand{\fez}{\phi^{(0)}}
\newcommand{\ftz}{\widetilde{\phi}^{(0)}}
\newcommand{\fbz}{\phi_{\scriptscriptstyle \rm B}}
\newcommand{\ftb}{{\widetilde{\phi}_{\scriptscriptstyle \rm B}}}
\newcommand{\hpz}{h_{\scriptscriptstyle \rm HW}^{+*}}
\newcommand{\Hjpm}{H_{{\rm j}}^{\pm}}
\newcommand{\Hupm}{H_{{\rm 1}}^{\pm}}
\newcommand{\fuz}{\phi_1^{(0)}}
\newcommand{\epsa}{\varepsilon_\alpha}
\newcommand{\epsg}{\varepsilon_\gamma}
\newcommand{\Oda}{{\cal O}_{{\rm exp}} (\epsa^2)}
\newcommand{\Lv}{L(\kzu x; \kzu b ; \dw)}
\newcommand{\Lvd}{L(\kzd x; \kzd b ; \dw)}
\newcommand{\Zv}{Z( \vq; \bt , \dw)}
\newcommand{\hpzv}{h_{\scriptscriptstyle \rm HW}^{+} }
\newcommand{\htpzv}{\tilde{h}_{\scriptscriptstyle \rm HW}^{+} }
\newcommand{\htphw}{\tilde{h}_{\scriptscriptstyle \rm HW}^{+} }
\newcommand{\Lz}{L}
\newcommand{\Lt}{L^{\scriptscriptstyle T}}
\newcommand{\Lhw}{L^{\scriptscriptstyle HW}}
\newcommand{\Oua}{{\mathcal O}\left( \epsa, \frac{\beta \ead}{b}\right)}
\newcommand{\oua}{{\mathcal o}\left( \epsa, \frac{\beta \ead}{b}\right)}
\newcommand{\fb}{\phi_{{\rm b}}}
\newcommand{\fuu}{\phi_1^{(1)}}
\newcommand{\Oeps}{{\mathcal O} (\eps)}
\newcommand{\Oexpeps}{{\mathcal O}_{{\rm exp}} (\eps)}
\newcommand{\Oexp}{{\mathcal O}_{{\rm exp}}}
\newcommand{\roawx}{\rho_{\alpha {\rm w}}(x)}
\newcommand{\cO}{{\cal O}}
\newcommand{\fdt}{\tilde{\phi}_2}
\newcommand{\vt}{{\bf t }}
\newcommand{\Frtt}{{\tilde F}_{\scriptscriptstyle{\rm RT}}}
\newcommand{\epsaap}{\varepsilon_{\alpha\gamma}}
\newcommand{\vu}{{\mathbf u}}
\newcommand{\wzpg}{w_0 (\xt '; \eps_\gamma, \dw)}
\newcommand{\wza}{w_0 (\xt ; \eps_\alpha, \dw)}
\newcommand{\ku}{\kappa_1}
\newcommand{\df}{\delta \phi}
\newcommand{\Ht}{\tilde{H}}
\newcommand{\OP}{\mathcal{O P}}
\newcommand{\OPU}{\mathcal{O P}_{{\rm U}}}
\newcommand{\OPUU}{\mathcal{O P}_{{\rm U}} \{ 1 \}(x) }
\newcommand{\OPUUn}{\mathcal{O P}_{{\rm U}} \{ 1 \} }
\newcommand{\OPUlinU}{\mathcal{O P}_{{\rm U_{\rm{lin}}}} \{ 1 \}(x)}
\newcommand{\OPUlinUn}{\mathcal{O P}_{{\rm U_{\rm{lin}}}} \{ 1 \}}
\newcommand{\OPUf}{\mathcal{O P}_{{\rm U}} \{ f \}(x) }
\newcommand{\ech}{\underset{\rightharpoondown}{\leftharpoonup} }
\newcommand{\Ei}{{\rm{Ei}}}
\newcommand{\Oc}{\mathcal{O}}
\newcommand{\psO}{\tilde{\mathcal{O}}}
\newcommand{\psOd}{\tilde{\mathcal{O}}\left( \eps^2 \right)}
\newcommand{\C}{\mathbf{C}}
\newcommand{\liml}{\underset{\eps \ll l \ll 1}{Lim}}
\newcommand{\Ulin}{U_{\rm{lin}}}
\newcommand{\Uuelin}{U_1^{* \, {\rm lin}}}
\newcommand{\Udelin}{U_2^{* \, {\rm lin}}}
\newcommand{\Uet}{\left( e^{\frac{\eps}{t}}-1 \right)}
\newcommand{\Uev}{\left( e^{\frac{\eps}{v}}-1 \right)}
\newcommand{\Lb}{\bar{L}}
\newcommand{\Ma}{M_\alpha}
\newcommand{\Mb}{\bar{M}}
\newcommand{\Mba}{{\bar{M}}_\alpha}
\newcommand{\epsd}{\eps_{{\scriptscriptstyle D}}}
\newcommand{\race}{\sqrt{|\eq|- \eta }\, \,  }
\newcommand{\hthw}{\tilde{h}_{{\scriptscriptstyle H \, W}}^{+}}
\newcommand{\Hjpe}{H_{{\rm j}}^{ + \, *}}
\newcommand{\hjep}{h_{{\rm j}}^{ + \, *}}
\newcommand{\Ujbmax}{{\bar U}_{\rm j}^{\rm max}}
\newcommand{\LUj}{ {\cal L}_{{\scriptscriptstyle U}_{\rm j }}}
\newcommand{\LUjlin}{ {\cal L}_{{\scriptscriptstyle U}_{\rm j
      }^{{\rm lin}}}}
\newcommand{\LUjbmax}{ {\cal L}_{{\scriptscriptstyle V}_{\rm
      j}^{\rm max}}}
\newcommand{\LmUjmax}{ {\cal L}_{-{\scriptscriptstyle U }_{\rm
      j}^{{\rm max}}}}
\newcommand{\LdifU}{ {\cal L}_{{\scriptscriptstyle U }_{\scriptscriptstyle
      j} - {\scriptscriptstyle U }_{\scriptscriptstyle
      j}^{{\rm  lin}}}}
\newcommand{\Lust}{ {\cal L}_{1/t}}
\newcommand{\Lgj}{ {\cal L}_{g_j}}
\newcommand{\HUj}{H_{{\scriptscriptstyle U}_{\rm j}}}
\newcommand{\HUjlin}{H_{{\scriptscriptstyle U}_{\rm j}^{{\rm lin}}}}
\newcommand{\RUj}{R_{{\scriptscriptstyle U}_{\rm j}}}
\newcommand{\RUjlin}{R_{{\scriptscriptstyle U}_{\rm j}^{{\rm lin}}}}
\newcommand{\HmUjmax}{H_{- {\scriptscriptstyle U}_{\rm j}^{{\rm max}}}}
\newcommand{\Uue}{U_1^*}
\newcommand{\Ude}{U_2^*}
\newcommand{\LU}{{\cal L}_{{\scriptscriptstyle U}}}
\newcommand{\LabsU}{{\cal L}_{\left| {\scriptscriptstyle U}\right|}}
\newcommand{\Labsg}{{\cal L}_{g}}
\newcommand{\Lmepsabsg}{{\cal L}_{- \eps g}}
\newcommand{\nth}{\mbox{n}^{\mbox{th}}}
\newcommand{\HU}{H_{{\scriptscriptstyle U}}}
\newcommand{\Hepsst}{H_{\eps / t}}
\newcommand{\Wepssdusd}{W_{\eps/2 \, , \, 1/2}}
\newcommand{\Uumax}{U_1^{{\rm max}}}
\newcommand{\Udmax}{U_2^{{\rm max}}}
\newcommand{\Ujmax}{U_{{\rm j}}^{{\rm max}}}
\newcommand{\Fj}{F_{{\rm j}}}
\newcommand{\epsj}{\eps_{{\rm j}}}
\newcommand{\uua}{u_{1 \, \alpha}}
\newcommand{\uda}{u_{2 \, \alpha}}
\newcommand{\Uulin}{U_1^{{\rm lin}}}
\newcommand{\hupe}{h_1^{+ *}}
\newcommand{\hume}{h_1^{- *}}
\newcommand{\hupme}{h_1^{\pm *}}
\newcommand{\hdpe}{h_2^{+ *}}
\newcommand{\Udlin}{U_2^{{\rm lin}}}
\newcommand{\roax}{\rho_\alpha (x)}
\newcommand{\eg}{e_\gamma}
\newcommand{\rog}{\rho_\gamma}
\newcommand{\hagrrp}{h_{\alpha \gamma} (\vr, \vr')}
\newcommand{\fcm}{\phi^{{\scriptscriptstyle CM}}}
\newcommand{\fcmlin}{\phi^{{\scriptscriptstyle CM, \,}{\rm lin}}_{z}}
\newcommand{\roacm}{\rho_\alpha^{{\scriptscriptstyle CM}}}
\newcommand{\ea}{e_\alpha}
\newcommand{\eap}{e_{\alpha'}}
\newcommand{\Vtotcm}{V_{{\rm tot}}^{{\scriptscriptstyle CM}}}
\newcommand{\dVtotcm}{\delta V_{{\rm tot}}^{{\scriptscriptstyle CM}}}
\newcommand{\zacm}{z_\alpha^{{\scriptscriptstyle CM}}}
\newcommand{\zgcm}{z_\gamma^{{\scriptscriptstyle CM}}}
\newcommand{\rocmlin}{\rho_\alpha^{{\scriptscriptstyle CM, \,}{\rm lin}}}
\newcommand{\Vtotcmlin}{V_{{\rm tot}}^{{\scriptscriptstyle CM, \,}{\rm
      lin}}}
\newcommand{\dVtotcmlin}{\delta V_{{\rm tot}}^{{\scriptscriptstyle CM, \,}{\rm
      lin}}}
\newcommand{\egd}{e_\gamma^2}
\newcommand{\Vtot}{V_{{\rm tot}}}
\newcommand{\Pb}{P^{{\rm B}}}
\newcommand{\kdb}{k_{{\scriptscriptstyle D}}}
\newcommand{\kb}{k_{{\scriptscriptstyle B}}}
\newcommand{\roahw}{\rho_\alpha^{{\scriptscriptstyle HW}}}
\newcommand{\Mbhw}{\bar{M}^{{\scriptscriptstyle HW}}}
\newcommand{\Lbhw}{\bar{L}^{{\scriptscriptstyle HW}}}
\newcommand{\Mhw}{M^{{\scriptscriptstyle HW}}}
\newcommand{\rogp}{\rho_{\gamma'}}
\newcommand{\egp}{e_{\gamma'}}
\newcommand{\Vtothw}{\Vtot^{{\scriptscriptstyle HW}}}
\newcommand{\rogb}{\rho_{\gamma}^ {\scriptscriptstyle{\rm B}}}
\newcommand{\Vas}{V_{{\rm as}}}
\newcommand{\Pas}{\Phi_{{\rm as}}}
\newcommand{\roaasym}{\rho_\alpha^{{ \rm asym}}}
\newcommand{\roasym}{\rho_\alpha^{{ \rm sym}}}
\newcommand{\equivx}{\underset{x \rightarrow + \infty}{\sim}}
\newcommand{\soma}{\sum_\alpha}
\newcommand{\rohw}{\rho^{{\scriptscriptstyle HW}}}
\newcommand{\Phw}{\Phi^{{\scriptscriptstyle HW}}}
\newcommand{\ropb}{\rho_{+}^ {{\rm B}}}
\newcommand{\sscm}{{\scriptscriptstyle CM}}
\newcommand{\Pcm}{\Phi^{\sscm}}
\newcommand{\Pcmlin}{\Phi^{\sscm, {\rm lin}}}
\newcommand{\rop}{\rho_+}
\newcommand{\rom}{\rho_-}
\newcommand{\romb}{\rho_{-}^ {{\rm B}}}
\newcommand{\kbe}{\kd \beta e^2}
\newcommand{\kdbf}{\kd b}\newcommand{\ei}{e_{{\rm i}}}
\newcommand{\ej}{e_{{\rm j}}}
\newcommand{\rj}{{\vr}_{{\rm j}}}
\newcommand{\ri}{{\vr}_{{\rm i}}}
\newcommand{\vc}{v_{{\rm c}}}
\newcommand{\vux}{{\mathbf u}_x}
\newcommand{\ssB}{{\scriptscriptstyle B}}
\newcommand{\roainf}{\rho_{\alpha}^{\ssB}}
\newcommand{\roarmu}{\rho_\alpha \left( \vr ; \{ \mu_\gamma \} \right)}
\newcommand{\QdD}{Q_{\delta \calD}}
\newcommand{\norme}{\vert {\mathbf e} \vert}
\newcommand{\vmu}{{\boldsymbol{\mu}}}
\newcommand{\vN}{{\mathbf N}}
\newcommand{\ve}{{\mathbf e}}
\newcommand{\cC}{\hat{C}}
\newcommand{\roaz}{\rho_\alpha (0)}
\newcommand{\emusepu}{\left( \frac{\ew - 1}{\ew + 1} \right)}
\newcommand{\emusepup}{\left( \frac{\varepsilon_w - 1}{\varepsilon_w +
      1} \right)}
\newcommand{\umesupe}{\left( \frac{1-\ew}{1+\ew} \right)}
\newcommand{\umesupep}{\left( \frac{1-\varepsilon_w}{1+\varepsilon_w}
  \right)}
\newcommand{\vsdu}{v_{{\scriptscriptstyle SD}}}
\newcommand{\vsdd}{V_{{\scriptscriptstyle SD}}}
\newcommand{\zux}{z^{[1]} (x)}
\newcommand{\zaux}{z_{\alpha}^{[1]} (x)}
\newcommand{\kudx}{\kappa_1^2 (x)}
\newcommand{\vl}{{\mathbf l}}
\newcommand{\kudinf}{(\kappa_1^{\ssB})^2 }
\newcommand{\kddx}{\kappa_2^2(x)}
\newcommand{\emur}{\epsilon_{{\scriptscriptstyle mur}}}
\newcommand{\Wax}{W_\alpha (x)}
\newcommand{\xid}{\xi_{{\scriptscriptstyle D}}}
\newcommand{\Gtd}{\Gamma^{3/2}}
\newcommand{\roinf}{\rho^{\ssB}}
\newcommand{\hA}{h_{{\scriptscriptstyle A}}}
\newcommand{\fireiriv}{v (\vr ; \{ e_i, \vr_i \}, V) } 
\newcommand{\Utelect}{U_{{\rm elect}}^*}
\newcommand{\muta}{\mu_\alpha^*}
\newcommand{\vp}{{\mathbf p}}
\newcommand{\laa}{\lambda_\alpha}
\newcommand{\laap}{\lambda_{\alpha'}}
\newcommand{\zta}{z_\alpha}
\newcommand{\ztg}{z_\gamma}
\newcommand{\cPi}{{\mathcal{P}}_i}
\newcommand{\somg}{\sum_\gamma}
\newcommand{\za}{z_\alpha}
\newcommand{\zg}{z_\gamma}
\newcommand{\vz}{{\mathbf 0}}
\newcommand{\Nb}{\bar{N}}
\newcommand{\Vsd}{\frac{V}{2}}
\newcommand{\bedsb}{\beta e^2 / b}
\newcommand{\sgegtrgb}{\somg \eg^3 \rogb}
\newcommand{\sgegdrgb}{\somg \eg^2 \rogb}
\newcommand{\scond}{\sigma_{{\rm cond}}}
\newcommand{\scondo}{\sigma_{{\rm cond}}^{{\rm o}}}
\newcommand{\scondd}{\sigma_{{\rm cond}}^{{\rm d}}}
\newcommand{\spropr}{\sigma_{{\rm propr}}}
\newcommand{\pa}{p_\alpha}
\newcommand{\pg}{p_\gamma}
\newcommand{\diagFcc}{\mbox{\begin{picture}(58,25)(13,-3)
\put(20,0){\circle{10}}
\put(25,0){\line(1,0){40}}
\put(65,0){\circle*{5}}
\put(40,10){$\Fcc$}
\put(18,10){$\scriptstyle{x}$}
\put(63,10){$\scriptstyle x'$}
\end{picture}}}
\newcommand{\diagFccq}{\mbox{\begin{picture}(58,25)(13,-3)
\put(20,0){\circle{10}}
\put(25,0){\line(1,0){40}}
\put(65,0){\circle*{5}}
\put(40,10){$\Fcc$}
\end{picture}}}
\newcommand{\diagFrt}{\mbox{\begin{picture}(58,25)(13,-3)
\put(20,0){\circle{10}}
\put(25,0){\line(1,0){40}}
\put(65,0){\circle*{5}}
\put(40,10){$\Frt$}
\end{picture}}}
\newcommand{\diagFmc}{\mbox{\begin{picture}(58,25)(13,-3)
\put(20,0){\circle{10}}
\put(65,0){\vector(-1,0){40}}
\put(65,0){\circle*{5}}
\put(40,10){$\Fmc$}
\end{picture}}}
\newcommand{\diagFcm}{\mbox{\begin{picture}(58,25)(13,-3)
\put(20,0){\circle{10}}
\put(25,0){\vector(1,0){38}}
\put(65,0){\circle*{5}}
\put(40,10){$\Fcm$}
\end{picture}}}
\newcommand{\diagfcc}{\mbox{\begin{picture}(65,25)(13,-3)
\put(20,0){\circle{10}}
\put(25,0){\line(1,0){40}}
\put(70,0){\circle{10}}
\put(40,10){$\fcc$}
\end{picture}}}
\newcommand{\diagfcm}{\mbox{\begin{picture}(65,25)(13,-3)
\put(20,0){\circle{10}}
\put(25,0){\vector(1,0){40}}
\put(70,0){\circle{10}}
\put(40,10){$\fcmq$}
\end{picture}}}
\newcommand{\diagfmc}{\mbox{\begin{picture}(65,25)(13,-3)
\put(20,0){\circle{10}}
\put(65,0){\vector(-1,0){40}}
\put(70,0){\circle{10}}
\put(40,10){$\fmcq$}
\end{picture}}}
\newcommand{\diagfmm}{\mbox{\begin{picture}(65,25)(13,-3)
\put(20,0){\circle{10}}
\put(65,0){\vector(-1,0){40}}
\put(25,0){\vector(1,0){40}}
\put(70,0){\circle{10}}
\put(40,10){$\fmm$}
\end{picture}}}
\newcommand{\diagdfcc}{\mbox{\begin{picture}(70,25)(13,-3)
\put(20,0){\circle{10}}
\put(25,0){\line(1,0){45}}
\put(50,0){\circle*{5}}
\put(75,0){\circle{10}}
\put(60,10){$\fcc$}
\put(30,10){$\fcc$}
\put(50,-15){$\zb$}
\end{picture}}}
\newcommand{\diagtfcc}{\mbox{\begin{picture}(70,25)(13,-3)
\put(20,0){\circle{10}}
\put(25,0){\line(1,0){45}}
\put(39,0){\circle*{5}}
\put(57,0){\circle*{5}}
\put(75,0){\circle{10}}
\put(27,10){$\fcc$}
\put(43,10){$\fcc$}
\put(60,10){$\fcc$}
\put(36,-15){$\zb$}
\put(56,-15){$\zb$}
\end{picture}}}
\newcommand{\zat}{z_\alpha}
\newcommand{\Phib}{\Phi^{{\rm B}}}
\newcommand{\vecz}{{\mathbf 0}}
\newcommand{\so}{\sigma^{{\rm o}}}
\newcommand{\sd}{\sigma^{{\rm d}}}
\newcommand{\rpbsestd}{\frac{\sqrt{\pi} \beta^{3/2}}{\esolv^{3/2}}}
\newcommand{\setrsredr}{\frac{\somg \eg^3 \rogb}{\left( \sum_\delta 
e_\delta^2 \rho_{\delta}^{{\rm B}} \right)^{1/2}}}
\newcommand{\dsetrsredr}{\somg \frac{ \eg^3 \rogb}{\left( \sum_\delta 
e_\delta^2 \rho_{\delta}^{{\rm B}} \right)^{1/2}}}
\newcommand{\sssc}{{\scriptscriptstyle sc}}
\newcommand{\Vselfcm}{\Vself^{{\sssc}}}
\newcommand{\Vselfcmlin}{\Vself^{{\sssc}, {\rm lin}}}
\newcommand{\rocp}{\rho_{{\scriptscriptstyle OCP}}}
\newcommand{\Mocp}{M_{{\scriptscriptstyle OCP}}}
\newcommand{\ssocp}{{\scriptscriptstyle OCP}}
\newcommand{\Sbulk}{S^{\ssB}}
\newcommand{\rob}{\rho^{\ssB}}
\newcommand{\vvp}{{\mathbf p}}
\newcommand{\ssW}{{\scriptscriptstyle W}}
\newcommand{\ewz}{\varepsilon_{\ssW_0}}
\newcommand{\vF}{{\mathbf F}}
\newcommand{\hH}{\hat{H}}
\newcommand{\hT}{\hat{T}}
\newcommand{\hV}{\hat{V}}
\newcommand{\hU}{\hat{U}}
\newcommand{\hvp}{\hat{\vp}}
\newcommand{\krrp}{\boldsymbol{|}\vr, \vr' \boldsymbol{>}}
\newcommand{\TrL}{{\rm Tr}_\Lambda}
\newcommand{\mua}{\mu_\alpha}
\newcommand{\Xig}{\boldsymbol{\Xi}}
\newcommand{\nap}{n_p^{\alpha}}
\newcommand{\etaa}{\eta_\alpha}
\newcommand{\roag}{\rho_{\alpha \, \gamma}^{(2)}}
\newcommand{\ssT}{{\scriptscriptstyle T}}
\newcommand{\roagT}{\rho_{\alpha \, \gamma}^{(2) \, \ssT}}
\newcommand{\hag}{h_{\alpha \, \gamma}}
\newcommand{\brarp}{\boldsymbol{<} \vr' \boldsymbol{|}}
\newcommand{\brar}{\boldsymbol{<} \vr \boldsymbol{|}}
\newcommand{\brarurd}{\boldsymbol{<} \vr_1, \vr_2 \boldsymbol{|}}
\newcommand{\brardru}{\boldsymbol{<} \vr_2, \vr_1 \boldsymbol{|}}
\newcommand{\braxp}{\boldsymbol{<} x' \boldsymbol{|}}
\newcommand{\ketr}{\boldsymbol{|} \vr \boldsymbol{>}}
\newcommand{\ketrurd}{\boldsymbol{|} \vr_1, \vr_2 \boldsymbol{>}}
\newcommand{\ketx}{\boldsymbol{|} x \boldsymbol{>}}
\newcommand{\hAop}{\hat{A}}
\newcommand{\hB}{\hat{B}}
\newcommand{\hUop}{\hat{U}}
\newcommand{\vxi}{\boldsymbol{\xi}}
\newcommand{\xix}{\xi_{{\rm x}}}
\newcommand{\xip}{\vxi_{\parallel}}
\newcommand{\md}{\frak{D}}
\newcommand{\Dx}{\frak{D}_{{\rm x}}}
\newcommand{\D}{\frak{D}}
\newcommand{\Erfc}{{\rm Erfc}}
\newcommand{\lsat}{\left( \lambda / a \right)^3}
\newcommand{\Lr}{\frak{L}}
\newcommand{\vwmvb}{\left[ \vw - \vb \right]}
\newcommand{\arxi}{\left( \alpha, \vr, \vxi \right)}
\newcommand{\rol}{\rho ( \Lr )}
\newcommand{\Lra}{\frak{L}_{{\rm a}}}
\newcommand{\Lrb}{\frak{L}_{{\rm b}}}
\newcommand{\vcc}{v^{{\rm c \, c }}}
\newcommand{\vcm}{v^{{\rm c \, m }}}
\newcommand{\vmc}{v^{{\rm m \, c }}}
\newcommand{\vmm}{v^{{\rm m \, m }}}
\newcommand{\bij}{\beta_{ i \, j}}
\newcommand{\zbl}{\zb ( \Lr )}
\newcommand{\fcmq}{f^{{\rm c \, m}}}
\newcommand{\fmcq}{f^{{\rm m \, c}}}
\newcommand{\fmm}{f^{{\rm m \, m}}}
\newcommand{\Lri}{\Lr_i}
\newcommand{\Lrj}{\Lr_j}
\newcommand{\Lrz}{\Lr_0}
\newcommand{\Lru}{\Lr_1}
\newcommand{\Lrd}{\Lr_2}
\newcommand{\ftt}{f_{\ssT \ssT}}
\newcommand{\Arz}{\frak{A}_0}
\newcommand{\fnc}{f^{- \, {\rm c}}}
\newcommand{\fcn}{f^{{\rm c} \, - }}
\newcommand{\Fnc}{F^{- \, {\rm c}}}
\newcommand{\Fcn}{F^{{\rm c} \, - }}
\newcommand{\kub}{\bar{\kappa}_1}
\newcommand{\dxdslad}{\frac{2 x^2}{\laa^2}}
\newcommand{\ssQ}{{\scriptscriptstyle Q}}
\newcommand{\ZQ}{Z^{\ssQ}}
\newcommand{\Lbq}{\bar{L}^{\ssQ}}
\newcommand{\kdl}{\kzd \lambda}
\newcommand{\lag}{\lambda_\gamma}
\newcommand{\lau}{\lambda_1}
\newcommand{\izuds}{\int_0^1 ds \,}
\newcommand{\nabg}{\boldsymbol{\nabla}}
\newcommand{\Mbq}{\Mb^{\ssQ}}
\newcommand{\Br}{\frak{B}}
\newcommand{\vf}{{\mathcal U}}
\newcommand{\plsapd}{\left( \frac{\lambda}{a} \right)}
\newcommand{\plsap}{\left( \lambda / a \right)}
\newcommand{\lsa}{ \frac{\lambda}{a}}
\newcommand{\kdebl}{\kd \lambda}
\newcommand{\Psia}{\Psi_{\alpha}}
\newcommand{\ma}{m_{\alpha}}
\newcommand{\UW}{U_{\scriptscriptstyle W}}
\newcommand{\vcW}{v_{\scriptscriptstyle CW}}
\newcommand{\vSR}{v_{\scriptscriptstyle SR}}
\newcommand{\eW}{\varepsilon_{\scriptscriptstyle W}}
\newcommand{\vcb}{v_{{\scriptscriptstyle C}{\rm bulk}}}
\newcommand{\Dpnu}{D_{\scriptscriptstyle W}^0}
\newcommand{\Dp}{{\overline D_{\scriptscriptstyle W}^0}}
\newcommand{\Dpt}{{\overline D_{\scriptscriptstyle W}}}
\newcommand{\DptD}{{\overline D_{\scriptscriptstyle DW}}}
\newcommand{\xiD}{\xi_{\scriptscriptstyle B}}
\newcommand{\alp}{\scriptscriptstyle\alpha}
\newcommand{\gam}{\scriptscriptstyle\gamma}
\newcommand{\rhoT}{\rho^{(2){\scriptscriptstyle T}}}
\newcommand{\phiW}{\phi_{\scriptscriptstyle W}}
\newcommand{\phiDW}{\phi_{\scriptscriptstyle DW}}
\newcommand{\phiDb}{\phi_{{\scriptscriptstyle D}{\rm bulk}}}
\newcommand{\kp}{{\bf k}_{\scriptscriptstyle //}}
\newcommand{\x}{{\widetilde x}}
\newcommand{\g}{{\scriptscriptstyle \gamma}}
\newcommand{{\hclW}}{h_{{\rm cl},{\scriptscriptstyle W}}}
\newcommand{{\hclDW}}{h_{{\rm cl},{\scriptscriptstyle DW}}}
\newcommand{\phiMFW}{\phi_{\scriptscriptstyle MFW}}
\newcommand{{\hclMFW}}{h_{{\rm cl},{\scriptscriptstyle MFW}}}
\newcommand{\rhodT}{\rho^{\scriptscriptstyle(2) T}}
\newcommand{\rhoW}{\rho_{\scriptscriptstyle W}}
\newcommand{\hW}{h_{\scriptscriptstyle W}}
\newcommand{\bzero}{\boldsymbol{0}}
\newcommand{\bnab}{\boldsymbol{\nabla}}
\newcommand{\bxi}{\boldsymbol{\xi}}
\def\a{{\scriptscriptstyle\alpha}}
\def\g{{\scriptscriptstyle\gamma}}
\def\d{{\scriptscriptstyle\delta}}
\def\y{\vy}
\newcommand{\vyt}{\tilde{\mathbf{y}}}
\newcommand{\felect}{f_{\rm elect}}
\newcommand{\fw}{f_{\omega}}
\newcommand{\fcl}{f_{\rm cl}}
\newcommand{\ssdd}{{\scriptscriptstyle 2D}}
\newcommand{\Gdd}{\Gamma_{\ssdd}}
\newcommand{\sun}{\sigma_1}
\newcommand{\sdeux}{\sigma_2}
\newcommand{\Ener}{\mathcal{E}}
\newcommand{\Erf}{{\rm Erf}}
\newcommand{\cnuag}{c_{{\rm nuage}}}
\newcommand{\Oetad}{\cO (\eta^2)}
\newcommand{\RLj}{R_{L_{{\rm j}}}}
\newcommand{\benote}{\begin{dinglist}{56}}
\newcommand{\ennote}{\end{dinglist}}
\newcommand{\calD}{{\mathcal D}}
\newcommand{\bgy}{{\textsc{bgy}} }
\newcommand{\ns}{n_{{\rm s}}}

\thispagestyle{empty}

\title{\bf Density profiles in a classical Coulomb fluid near a
  dielectric wall. II. Weak-coupling systematic expansions}
\author{Jean-No\"el AQUA and Fran\c{c}oise CORNU \\
Laboratoire de Physique Th\'eorique \thanks{Laboratoire associ\'e
au Centre National de la Recherche Scientifique - UMR 8627} \\
B\^atiment 210, Universit\'e Paris-Sud\\ 
91405 Orsay Cedex, France}

\maketitle
\begin{abstract}
  In the framework of the grand-canonical ensemble of statistical mechanics, we give an 
exact diagrammatic representation  of the density profiles in a classical multicomponent plasma near a dielectric
wall. By a reorganization of Mayer
diagrams for the fugacity expansions of the densities, we exhibit how
the long-range of both the self-energy and pair interaction are
exponentially screened at large distances from the wall. 
However,  
 the self-energy due to Coulomb interaction with images
 still diverges in the vicinity of the dielectric
wall and the variation of the density is drastically different at short 
or large distances from the wall. This variation is involved in the
inhomogeneous Debye-H\"uckel equation obeyed by the screened pair potential.
Then the main difficulty lies in the
determination of the latter potential at every distance. We solve
this problem by devising a systematic expansion with respect
to the ratio of the fundamental length scales involved in the two
coulombic effects at stake. (The application of this method to a plasma confined
between two
ideally conducting plates and to a quantum plasma will be presented
elsewhere). As a result we derive the exact analytical perturbative expressions for the density 
profiles up to first order in the coupling between charges. The mean-field
 approach displayed in Paper I is then justified.

{\bf KEYWORDS~:} Coulomb interactions, dielectric wall, grand-ca\-no\-ni\-cal ensemble, systematic 
resummations, inhomogeneous Debye-H\"uckel equa\-tion, screened potential with two 
characteristic length sca\-les.

\vskip 1cm 
\end{abstract}

\newpage

\numberwithin{equation}{section}

\section{Introduction}

The present paper is devoted to the  systematic derivation of the density 
profiles which are discussed in Paper I. The 
system is a classical multicomponent plasma, made of at least two 
species of moving charges
 with opposite signs, near a plane wall macroscopically characterized by a dielectric 
constant. The exact analytic expressions are obtained in the limit of 
weak Coulomb coupling  inside the fluid.

Our calculations
performed in the framework of statistical mechanics cast a new light 
on the fundamental phenomenon in such systems : Coulomb screening of
surface polarization charges inside the fluid. Indeed, from
electrostatics  it is
well known that the bulk thermodynamic properties in a fluid of charges
are independent from the charge state of boundary
walls. Nevertheless the microscopic long-ranged Coulomb pair interaction
$v(\vr ;\vrp)$ between two unit charges located respectively at $\vr$
and $\vrp$ takes into account the electrostatic response of
the wall~: $v(\vr ;\vrp)$ is a solution of Poisson equation 
\begin{equation}
  \Delta_{\vrp} v(\vr;\vrp) = - 4 \pi \delta(\vr - \vrp)
 \label{pois}
\end{equation}
and its expression is ruled  by the electrostatic boundary
conditions. For instance, when the material of the wall is
characterized by a relative dielectric constant
$\ew$ (with respect to the dielectric constant in the vacuum), the
solution of \eqref{pois} reads 
\begin{equation}
\vvw = \frac{1}{|\vr - \vrp|} - \dw \frac{1}{|\vr - \vrp ^*|}
\label{defvw}
\end{equation}
where $\dw \equiv (\ew-1)/(\ew+1) $ and $\vrp ^*$ is the image of
$\vrp$ with respect to the plane interface \cite{Jackson}. On the contrary, far away 
from any boundary or when the wall has no electrostatic response ($\ew =1$) the potential 
takes its ``bulk'' value, 
\begin{equation}
  \label{}
  \vb(\vr;\vrp) = \frac{1}{|\vr-\vrp|}
\end{equation}
Short-distance cut-offs must be introduced
in order to prevent the collapse between charges with opposite 
signs or between every charge and its image when $\ew >1$. 

At the inverse temperature 
$\beta = 1/\kb T $, where $\kb$ is Boltzmann constant,  
the four length scales in the problem are : the closest approach
distance $\beta e^2$ determined by Coulomb interaction between charges
of typical magnitude $e$ and with a mean kinetic energy of order
$1/\beta$; the mean interparticle distance $a$; the range $b$
of the hard-core repulsion from the wall (which is chosen to be the same
for all species); the hard-core diameter $\sigma$ of charges 
which prevents the electrostatic collapse inside the fluid. However the
cut-off $\sigma$ proves not to arise in the densities and correlations
at leading order in a low-density regime (because the long range of
Coulomb interactions is then the most important effect).

In the grand canonical ensemble (Section \ref{section2}) 
the macroscopic parameters are the volume of the system, the 
fugacities $z_\alpha$'s,  where $\alpha$ is a species
index running from 1 to the number of species $n_s$, 
and the inverse temperature $\beta$. The task of getting exact analytic 
results for a multicomponent plasma 
where a cut-off is added, and where the screening of the long-range part
of the interaction is systematically dealt with, was achieved for the
density expansion of bulk
correlations by Meeron \cite{Meer58}. Haga extended Meeron's scheme to
the low-density expansion of the pressure \cite{Haga} in the bulk where
densities are uniform\footnote{Apart from a minor error in the 
calculation of the coefficient called $S_2$, his result is correct.}.

In Section \ref{section3}
we devise a new reorganization of Mayer
fugacity-expan\-sions for every particle density 
$\roa(x;\{z_\gamma\},\beta)$ 
and its uniform bulk limit $\roab(\{ z_\gamma \},$ $\beta)$ 
in order to exactly deal with the long range of the self-energy of a charge
interacting with its image inside the wall as well as with the long
range of Coulomb pair interactions. ($x$ is the distance from the plane wall
located at $x=0$. The notations $\{ z_\gamma \}$
means that the fugacities of all species are involved.)
Our reorganization of diagrams is performed in two steps. In the first
step we exhibit how the screening of the bare one-body interaction with
the dielectric wall may be described by resummations of ``ring''
subdiagrams. In the second step, we use an already known method in order
to exactly handle with the screening of Coulomb pair interactions
\cite{Cornu96I}. (We notice that, if we had used a one-step resummation analogous
to that performed in Ref.\cite{ACP94} or \cite{Cornu96I}, there would have been resummed diagramms with
some bare  self-energies left -- and therefore some spurious $1/x$ tails left--
which would disappear only by an adequate gathering with other diagrams to be
specified in every particular case.)

The latter exact systematic resummations produce
auxiliary screened potentials $\fu$ and $\fd$ each of which obeys
a second-order differential equation studied in Section \ref{section4}.
The technical difficulty lies in the resolution of these auxiliary 
inhomogeneous Debye-H\"uckel equations.

In order to get explicit analytic expressions, we consider a regime of weak coupling inside the fluid
\begin{equation}
  \label{X}
  \Gamma \equiv \frac{ \beta e^2}{a} \ll 1.
\end{equation}
However, the electrostatic coupling with the dielectric wall is not 
necessarily weak even in the very vicinity of the wall,
\begin{equation}
  \label{}
  \frac{\beta e^2}{b} \sim 1
\end{equation}
 After resummation of coulombic
long-range divergencies, collective screening effects operate on a scale
equal to the bulk Debye length, $\kd^{-1}$, with 
\begin{equation}
  \label{defkdbis}
  \kd \, \equiv \, \sqrt{4 \pi \beta \sum_\alpha \ead \roab}
  \, \propto \,  \frac{\Gamma^{\frac{1}{2}}}{a} 
\end{equation}
The weak-coupling condition \eqref{X} is equivalent to the condition  
\begin{equation}
  \label{}
   a^2 \ll \kd^{-2}
\end{equation}
At the same time,  the density profile proves to be an expansion in integer
powers -- possibly multiplied by some powers of logarithms -- 
 not of the coupling parameter
$\beta e^2 / a = \Gamma $ but of
\begin{equation}
  \label{defepsilon}
  \epsd \, \equiv \, \frac{1}{2} \beta e^2 \kd \, \propto \,
  \Gamma^{3/2} \ll 1
\end{equation}
Thus an $\epsd$-expansion provides an expansion in powers of the
square root of the density. (The cut-off $\sigma$ for pair interactions
corresponds to an integrable interaction ; so it
arises in the  corrections to bulk
quantities only from the relative order proportional to the density, namely 
from the  order $\epsd^2$ .)

We show that the screened potentials 
$\fu$ and $\fd$ can be expanded
in powers of the ratio $\tilde{\lambda}$ of the two length scales $\beta e^2$
and $\kd^{-1}$ which   controls the variations of the $x$-dependent inverse
Debye length 
in the equation obeyed by each $\phi_i$. The expansion of each $\phi_i(\vr,\vr')$ 
is uniform with respect
to the  distances $x$ and $x'$ from the wall -- when the projection of
$\vr-\vr'$  onto the wall is kept fixed --  in the sense that the  term of 
order
$\tilde{\lambda}^n$ is  bounded for all $x>b$ and $x'>b$  by a constant which
depends on
$\kd b$ and $\tilde{\lambda}/[\kd b]=\beta e^2/b$. In fact $\tilde{\lambda}$ 
coincides with
$\epsd$ in the present problem and  
 a scaling analysis of Mayer diagrams allows
one to obtain the small-$\epsd$ expansion of the density profile (see
Section \ref{section5}).
Only one diagram
contributes up to  first  order in $\epsd$
and $\roa ( x ; \{ z_\gamma \}, \beta ) $ can be explicitly rewritten as 
$\roa ( x ; \{ \rogb \}, \beta)$. 
The latter general scheme, which provides systematic expansions in
$\tilde{\lambda}$ and $\epsd$, can be also applied to a classical plasma near a
charged dielectric or an ideally conducting wall or to a low-degenerated
quantum plasma in the vicinity of a boundary with any electrostatic
response. The corresponding works will be displayed elsewhere.

Eventually, at first order in the coupling parameter $\epsd$, 
\begin{equation}
  \label{structrho}
  \roax = \roab \, \theta (x-b) \, \exp 
 \left[ \dw \beta \ead \frac{e^{-2  \kd x}}{4x} \right]
 \times \left[ 1 + R(\kd x ; \epsd, \kd b) \right]
\end{equation}
where $R$ is of order $\epsd \ln \epsd$ and $\epsd$.
$R(\kd x)$ decays exponentially fast at large
distances over the scale $\kd^{-1}$ and tends to a finite value when $x$
comes to zero. Subsequently,
$\roax$
varies drastically over the bare-coupling scale $\beta e^2$ very close
to the wall, whereas its variation at distances larger than $\kd ^{-1}$
is scaled by the screening length $\kd^{-1}$. In other words, very near
the wall, $\roax$ is governed by the part of the self-energy of each particle
originating from the electrostatic response of the dielectric wall,
whereas the mean Coulomb interaction  with other charges of the fluid
modified by the impenetrable wall
dominates the approach to the bulk value far away from the wall. 
After calculation of the electrostatic potential drop $\Phi(x)$ at leading order
$\epsd$ from the charge 
density profiles given by \eqref{structrho}, the density profiles prove to coincide 
with the expressions derived from the mean-field approach introduced in 
Paper I, 
\begin{equation}
  \label{}
    \roax = \roab \, \theta (x-b) \, e^{- \beta \ead \Vselfsc (x) } \left[ 1 -
  \beta \ea \Phi(x) \right]
\end{equation}
where $\Vselfsc(x)$ is the screened self-energy at order $\epsd$ arising from
the first step in the resummations of Mayer diagrams.


\section{Model}
\label{section2}

\subsection{Potential energy}

The total potential energy is the sum of the electrostatic energy $\Uelec$
and the short-range repulsive interaction $\usr$. $\Uelec$ is equal to the
sum of Coulomb pair interactions plus the sum of self-energies
in the presence of the wall,
\be
\Uelec = \frac{1}{2} \sum_{i \neq j} e_{\alpha_i} e_{\alpha_j}
\vw (\vr_i; \vr_j) + \sum_i e_{\alpha_i}^2 V_{{\rm self}} (x_i)
\label{defUelect}
\end{equation}
In \eqref{defUelect} $e_{\alpha_i}$ is the charge of particle i and
\be
V_{{\rm self}} (x_i) = -\dw \frac{1}{4 x_i}
\label{defVself}
\end{equation}
where $\dw$ is defined in \eqref{defvw}.
(see Ref. \cite{Land&lifs8}). 
The repulsion energy $\usr$ is also the sum of hard-core pair interactions plus
the sum of one-body potentials created by the wall,
\be
\label{cddefusr}
\usr = \frac{1}{2} \sum_{i \neq j} \vsr (\vr_i - \vr_j) + \sum_i
\Vsr (x_i) 
\end{equation}
with 
\begin{equation}
\exp \left[ - \beta \vsr (\vr_i; \vr_j) \right] = 
\left\{
\begin{array}{ll}
0 & \mbox{if $|\vr_i - \vr_j|<\sigma$}\\
1 & \mbox{if $|\vr_i - \vr_j|>\sigma$}
\end{array}
\right.
\end{equation}
and
\begin{equation}
\exp \left[ - \beta \Vsr (x) \right] = 
\left\{
\begin{array}{ll}
0 & \mbox{if $x<b$} \\
1 & \mbox{if $x>b$}
\end{array}
\right.
\label{defVsr}
\end{equation}
We notice that in the case $\ew <1$, $\exp \left( - \beta \ead \Vself
  (x) \right)$ vanishes when $x$ goes to zero and $\Vsr$ can be omitted,
  i.e. $b$ can be set to zero, without any collapse of the system onto the wall.


\subsection{Fugacity expansions}

The grand partition function of the system in the semi-infinite space
$x>0$ reads
\begin{equation}
\label{part}
{\cal \Xi} ( \beta, \{ z_\gamma \} ) = \sum_{N=0}^{+\infty} \frac{1}{N!} \int
\left[ \prod_{i=1}^{N} d {\cal P}_i \, z_{\alpha_i} \right] e^{-\beta
\left(\Uelec + \usr \right)} 
\end{equation}
In \eqref{part} the integration measure is denoted by $\int d{\cal P}_i \equiv
\displaystyle\int_{x_i>0} d\vr_i \displaystyle\sum_{\alpha_i = 1}^{n_s}$
and ${\cal P}_i$ is the notation for $(\vr_i,\alpha_i)$. $z_{\alpha_i}$
has the dimension of an inverse cubed distance. The one-body
interactions can be absorbed in an $x$-dependent fugacity
\begin{equation}
 \label{defzb}
\zb_{\alpha_i} (x_i) \equiv z_{\alpha_i} e^{-\beta \left[ \Vsr
  (x_i) + e_{\alpha_i}^2 \Vself (x_i) \right]}
\end{equation}
According to \eqref{defVself} and \eqref{defVsr}
\be
\zb_{\alpha_i} (x_i) = z_{\alpha_i} \theta (x_i - b) e^{\dw \beta
  e_{\alpha_i}^2/(4x_i)}
\label{ztheta}
\end{equation}
where $\theta$ is the Heaviside function. $\Xi$ can be written as 
\be 
\Xi ( \beta, \{ z_\gamma \} ) = \sum_{N=0}^{+\infty} \frac{1}{N!} \int
\left[ \prod_{i=1}^{N} d {\cal P}_i \, \zb_{\alpha_i}(x_i) \right] 
e^{-\frac{\beta}{2}\sum_{i \neq j}  \left[ v_{\rm SR}(\vr_i - \vr_j) +
    e_{\alpha_i} e_{\alpha_j} \vw (\vr_i ; \vr_j)  \right]} 
\label{xi}
\end{equation}
The particle density for species $\alpha$ is derived from the
grand partition function through a functional derivative
\be
\roaw  = \zb_\alpha (x)
  \left. \frac{\delta \ln \Xi \left( \beta, \{ z_\gamma^*\} \right)}{\delta
    z_\alpha^*(x)} \right|_{z_\alpha^* (x) = \zb_\alpha (x)}
\label{relrhoxi}
\end{equation}

By definition of the bulk 
\begin{equation}
\label{defrhob}
\lim_{x \rightarrow + \infty } \roaw= \roab (\beta, \{ 
z_\gamma \} )
\end{equation}
According to the survey of the existence of thermodynamic limit for
Coulomb fluids \cite{Lieb&Lebo72}, the bulk densities satisfy the local
neutrality
\be
\sum_\alpha e_\alpha \roab (\beta, \{ z_\gamma \} ) = 0 
\label{localneutr}
\end{equation}
whatever the values of the $z_\gamma$'s are. In the following we will
consider a weak coupling regime where the system behaves as an ideal gas
at zeroth order in $\Gamma$. Besides, the particle
density of species $\alpha$ in an ideal gas is just proportional to the 
fugacity $z_\alpha$. As a consequence, we can take advantage of the
freedom 
of choice for the $z_\gamma$'s exhibited in Ref. \cite{Lieb&Lebo72} and
enforce the extra condition 
\be 
\sum_\gamma e_\gamma z_\gamma = 0
\label{zneut}
\end{equation}
\eqref{zneut} ensures that the ideal gas also satisfies
\eqref{localneutr} and this provides a convenient simplification in the
relation between the $\rho_\gamma$'s and the $z_\gamma$'s. 
We stress that the  irrelevance of the value taken by one among the
fugacities, which is always valid  for bulk quantities,  also 
holds for surface statistical properties 
but only when the wall gets no global influence
charge in the
presence of the Coulomb fluid. This is indeed the case when $\ew$ is
finite. On the contrary, when $\ew$ is sent to infinity, the wall
material is an ideal conductor which gains a global charge by influence. 
The latter
is exactly compensated by the net surface charge in the fluid and it
depends on all fugacities $z_\gamma$'s.


\subsection{Representation of the density profile in terms of Mayer diagrams}

The fugacity-expansion \eqref{xi} may be expressed 
in terms of Mayer diagrams where a 
point is associated with every $\cp_i$ and two points are joined by at
most one bond 
\begin{equation}
f(\cp_i,\cp_j) = e^{-\beta \left( \vsr(\vr_i - \vr_j)
  + e_{\alpha_i} e_{\alpha_j} \vw (\vr_i ; \vr_j) \right) } - 1  
\label{deff}
\end{equation}
The relation \eqref{relrhoxi} and topological arguments
\cite{HansenMacDo} lead to various diagrammatic representations for the
fugacity expansion of $\rho_{\alpha} (x)$. In the following we
  will use
\begin{equation}
\roa (x) = \zb_\alpha (x) \, \, \exp \left[ \sum_{\cg}
\frac{1}{S_{\cg}} \int \prod_{n=1}^{N} \left[ d{\cp}_n \zb_{\alpha_n} (x_n)
\right] \, \, \left[ \prod f \right]_{{\cg}} \right]
\label{diagrho}
\end{equation}
In \eqref{diagrho} the sum runs over all the unlabeled and
topologically different connected diagrams ${\cg}$ with one root
point ${\cp} \equiv (\alpha, \vr)$ (which is not integrated over) and
$N$ internal points ($N=1,\cdots,\infty$). Moreover the ${\cg}$'s
satisfy the extra constraint that the root point
$\cp$ is not an articulation point. (${\cp}$ is an
articulation point of ${\cg}$ if  ${\cg}$ is split into at least
two pieces when the point ${\cp}$ is removed.)
$\left[ \prod f \right]_{{\cg}}$ is the product of the $f$ bonds in
the diagram ${\cg}$ and $S_{{\cg}}$ is the symmetry factor of
${\cg}$, namely the number of permutations between some internal points
${\cp}_n$ that do not change the product $\left[ 
\prod f \right]_{\cg}$. 

At large distances $\vw (\vr_i ; \vr_j)$ behaves essentially as $1/
|\vr_i - \vr_j|$ and the $f$ bonds \eqref{deff} are not
integrable. The divergences in Mayer diagrams ${\cg}$ may be removed
by splitting the $f$ bonds into several auxiliary bonds $f^*$ and by
performing a reorganization of the diagrams ${\cg}^*$ made with $f^*$
bonds. ( $\cg^*$ diagrams have the same topological properties as
$\cg$ diagrams.) The decomposition of $f$ chosen for our purpose is
\begin{equation}
\label{cdfraclien}
f= \fcc + \frac{1}{2} \left[ \fcc \right]^2 + \ft 
\end{equation}
with
\be
\fcc (\cp_i,  \cp_j) \equiv - \beta e_{\alpha_i} e_{\alpha_j} \vw
(\vr_i;\vr_j)
\label{deffcc}
\end{equation}
With these definitions $\ft$ is at the border of integrability. The
systematic partial resummations of diagrams ${\cg}^*$
are described in next section.


\section{Exact topological resummations valid for any density}
\label{section3}

\subsection{Step 1 : Screening of the self-energy induced by the wall}
\label{cdpremresom}

The long-range one-body interaction $e_\alpha^2 \Vself (x) $ involved
in the fugacity $\zb_\alpha (x)$ turns into a short-range effective
interaction when Coulomb ``ring'' subdiagrams are resummed. We introduce
the following definitions. A Coulomb ``ring'' subdiagram carried by a
point $\cp$ is either a bond $\left[ f^{cc} \right]^2 /2 $ or a chain of
bonds $f^{cc}$ whose both ends coincide with $\cp$ and which contains at
least two intermediate points. 
The value $\Ir$ of the sum of all rings attached to a point $\cp$ is equal to
\be
\Ir(x) = - \frac{1}{2} \beta e_\alpha^2 \left[ \left( \fu - \vw
  \right) (\vr;\vr) \right] 
\end{equation}
where $-\beta e_\alpha^2 \fu(\vr;\vr')$ is the sum of chains made of
an arbitrary number of bonds
$\fcc$ between $\vr$ and $\vr'$ and with intermediate points $\cp_i$
weighted by
$\zb ( \cp_i ) $. ($1/2$ is the
symmetry factor of a single ring diagram.) By definition
\begin{multline}
-\beta e_\alpha e_{\alpha'} \, \fu (\vr;\vr') = f^{cc} (\cp;\cp')
  + \\
 \sum_{N=1}^{+\infty} \int \left[ \prod_{i=1}^{N} d\cp_i \, \zb (\cp_i) \right]
f^{cc}(\cp;\cp_1) f^{cc}(\cp_1;\cp_2) \cdots f^{cc}(\cp_N;\cp')
\label{defphi1}
\end{multline}
According to \eqref{deffcc}, the definition \eqref{defphi1} of $\fu$ 
is equivalent to the integral relation 
\begin{equation}
\fu(\vr;\vr')=\vw(\vr;\vr') - \beta  \sum_\alpha e_\alpha^2 \int d\vr''
\zb_\alpha (x'') \vw(\vr;\vr'') \fu(\vr'';\vr')
\label{intrelphi1}
\end{equation}
The sum of an arbitrary number of rings carried by a point $\cp$ is
equal to $\exp \Ir (\vr)$. (An analogous calculation appears in section
V.B of Ref.\cite{ACP94}.)

\begin{figure}
\centering\epsfig{figure=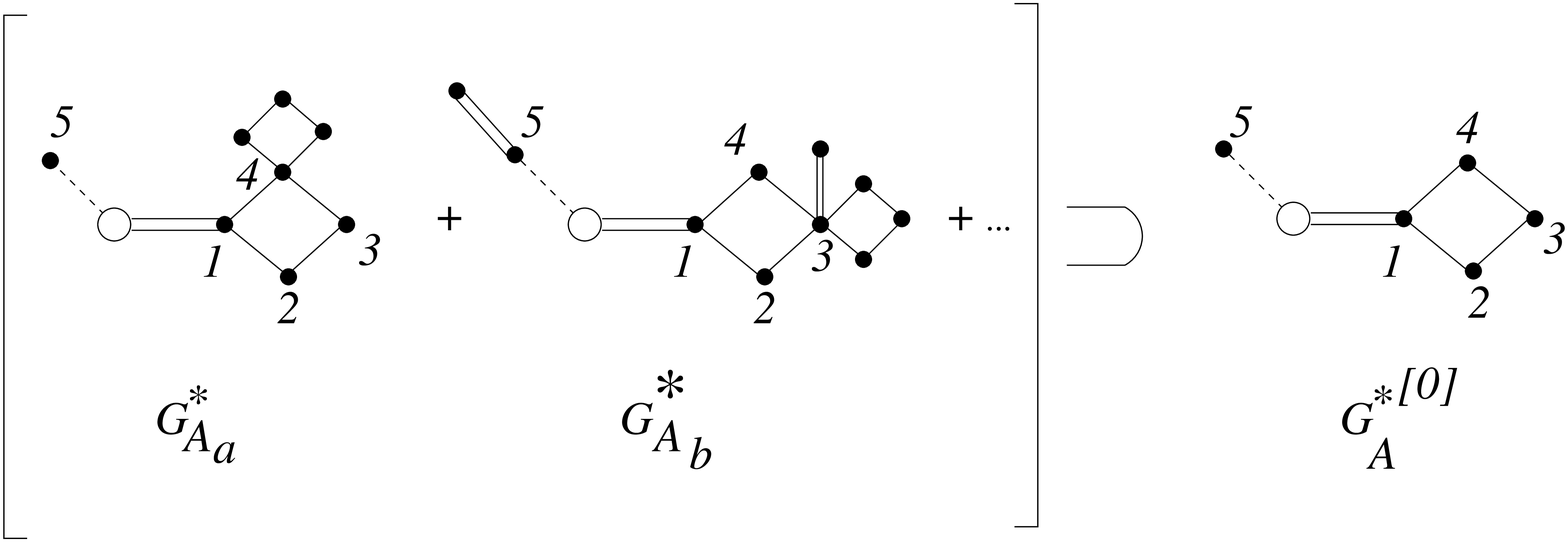, width=12cm}
\caption{Step-1 for resummations. $G_{A\, a}^*$ and $G_{A \, b}^*$ are 
two diagrams that only differ from $G_A^{* \, [0]}$ by ring
subdiagrams. Labels are attached to points which are common to all
diagrams. Each point $\cp_i$ carries a weight 
$\zb \left(\cp_i\right)$. A bond $\fcc$ is drawn as a solid line, a bond 
$(1/2)[\fcc]^2$ as a double solid line and a bond $\ft$ as a
dashed line. The subdiagram $\cs \left(\cp_1 \right)$ made of points
$\cp_2, \cp_3, \cp_4$ exists in $G_A^{* \, [0]}$ of Fig. \ref{f1}
because at least one of the latter points carries a ring in diagrams
$G_{A \, a}^*$ and $G_{A \, b}^*$.
} \label{f1}
\end{figure}

\begin{figure}
\centering\epsfig{figure=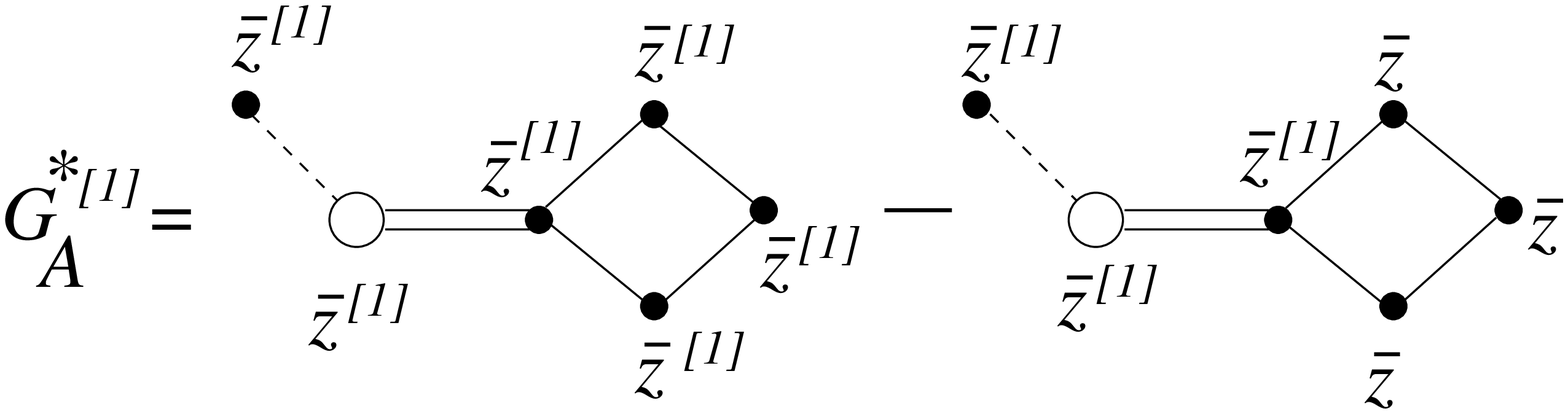, width=12cm}
\caption{Example for rule $\rr$ after step-1 resummations. After ring
  summations, $G_{A \, a}^*$ and $G_{A \, b}^*$ contribute to 
$G_A^{* \,  [1]}$, whereas the subdiagram $\cs \left( \cp_1 \right)$ in
 $G_A^{* \,  [0]}$ disappears and $G_A^{* \, [0]}$ contributes to another
 $G^{* \,  [1]}$. As a result the value of $G_A^{* \, [1]}$ is equal to 
  an integral where all bonds are the same as in 
$G_A^{* \,  [0]}$ but where all weights $\zb $ have been
  replaced by $\zb^{[1]} $ minus the value of the
  integral corresponding to $G_A^{* \, [0]}$ where the  weights 
 $\zb $ have been replaced by $\zb^{[1]} $ only for points which are not 
 the  intermediate
 points 
 of a ring subdiagram.
} \label{f2}
\end{figure}

Let us consider all
diagrams $\cg^*$ (with weight $\zb$) that only differ from a given
diagram $\cg^{* [0]}$ by one or more Coulomb ring subdiagrams 
carried by at least one point of $\cg^{* [0]}$ (see Fig.\ref{f1}).
Before integration over the internal points of $\cg^{* [0]}$, 
the integral corresponding to such a diagram is equal to the 
integrand associated with $\cg^{*  [0]}$ times the contributions 
from Coulomb rings. As a consequence, if we sum all such 
diagrams $\cg^*$  we get one diagram $\cg^{* [1]}$ which 
has the same points and bonds as $\cg^{*  [0]}$  except that 
the weights $\zb$ are replaced by weights
\begin{equation}
\zb^{[1]} (\cp) = \zb (\cp) \exp \left[ - \frac{1}{2} \beta e_\alpha^2
  (\fu - v_{\rm w}) (\vr;\vr) \right]
\label{defz1}
\end{equation}
where $\fu(\vr;\vr')$ is defined in \eqref{defphi1}. 
At the end of the first partial resummation, the diagrams
$\cg^*$ with bonds $\fcc$, $1/2 \, \left[ \fcc \right]^2$ and $\ft$ and
with weights $\zb_\alpha (x)$ have been replaced by 
diagrams  $\cg^{*  [1]}$ with the same kind of bonds but with weights
 $\zbau (x)$  and an extra construction rule $\rr$ (see Fig.\ref{f2}). 
 The latter one is 
necessary to avoid double counting. If some subdiagram $\cs(\cp_m)$ of
$\cg^{* [1]}$ is a ring of bonds $\fcc$ with weights $\zbau$
carried by the point $\cp_m$ (with weight  $\zb_{\alpha_m}^{[1]} (x_m)$), 
then after integration over intermediate points in the ring, the contribution 
from $\cs (\cp_m)$, which depends only on $x_m$, is factorized in the total 
contribution from $\cg^{* [1]}$ and is multiplied by the weight
$\zb_{\alpha_m}^{[1]} (x_m)$ of its root point $\cp_m$; rule $\rr$ states that 
the contribution from $\cs (\cp_m)$ is equal to its value where all its 
intermediate points have a weight $\zbau$ minus the
corresponding value where all its intermediate points would have a weight
$\zb_\alpha$. Indeed the latter value is already taken into account in
the weight $\zb_{\alpha_m}^{[1]} (x_m)$ of the point $\cp_m$, whereas a
subdiagram of $\cg^{*}$ made of bonds $\fcc$ but where at least one
intermediate point carries another Coulomb ring does not disappear in
the ring resummation process leading to $\cg^{* [1]}$. 
(See Figs.\ref{f1} and \ref{f2}).

According to the definition \eqref{defzb} of $\zb(\cp)$ and since
 $\Vself (x) = (1/2) \left( \vw \right. $ 
$\left. - \vb \right) (\vr;\vr)$, \eqref{defz1} may be written as 
\be
\zbau (x) = \theta (x-b) z_\alpha \exp \left[ - \frac{1}{2} \beta
  e_\alpha^2 \left( \fu-\vb \right) (\vr;\vr) \right]
\label{valuez1}
\end{equation}
As shown below in a weak-coupling expansion, we expect that for any 
finite coupling constant $(\fu-\vb)(\vr;\vr)$ has no algebraic tail at large
distances. Thus, contrary to $\zb_\alpha (x)$ the screened fugacity
$\zbau (x)$ does not contain the long-range part of the
self-energy induced by the electrostatic response of the wall when 
$\ew \neq 1$. In other words, the resummation of rings subdiagrams has captured
the main effect of Coulomb screening upon the electrostatic interaction with the
wall.


\subsection{Step 2 : Screening of  pair interactions}

In the second step we sum all chains of bonds $\fcc$ in diagrams $\cg^{*
  [1]}$. The sys\-tematic resummation process is performed as follows. We
define a ``Cou\-lomb'' point as the intermediate point of a chain of two
bonds $\fcc$ and which is not linked to any other point in the
diagram. Such a point has a weight $\zbau$. A so-called
``prototype'' diagram $\cpp$ is a diagram $\cg^{* [1]}$ which
contains no Coulomb point. We sum all diagrams $\cg^{*[1]}$ which can be
  built from the same diagram $\cpp$ by addition of at least one
  Coulomb point (with weight $\zb^{[1]}$) with the associated $\fcc$
  bonds either to replace a $\fcc$ bond in $\cpp$ by  a chain of $\fcc$ bonds  or to
  multiply a bond in $\cpp$ (see Fig.3). In the latter case, we use the convention
  according to which the diagram $\cpp$ is the same whether the diagrams
  $\cg^{*[1]}$ contain a bond 
$(1/2) \left[ \fcc \right]^2 (\cp_i,  \cp_j )$ or a bond $\ft (\cp_i, \cp_j)$
which are possibly 
  multiplied by a chain of $\fcc$ bonds between $\cp_i$ and $\cp_j$.
Since the $\cpp$ diagram may contain articulation points,
  the resummation produces two effects. 

First, we sum all diagrams $\cg^{* [1]}$ that differ from $\cpp$
by a ring of bonds $\fcc$ with at least one intermediate point with
weight $\zb^{[1]}$. As in the reorganization of diagrams performed in
step 1, this process leads to a renormalization of the fugacity by an
exponential factor analogous to \eqref{defz1} with an extra 
subtraction arising
from the construction rule $\rr$ for $\cg^{*[1]}$ diagrams. The resummed
fugacity reads
\be
\zbad (x) = \zbau (x) \exp \left\{ -\frac{1}{2} \beta e_\alpha^2 \left[ \left
      ( \fd - \vw \right) (\vr;\vr) - \left(\fu - \vw
    \right)(\vr;\vr) \right] \right\}
\end{equation}
where $\fd$ is defined by the integral equation \eqref{intrelphi1} with
$\zb_\alpha (x)$ replaced by $\zbau (x)$. According to \eqref{defz1} and 
\eqref{ztheta}, the expression of $\zbad (x)$ is similar to
\eqref{valuez1} 
\be
\label{valuez2}
\zbad (x) = \theta (x-b) z_\alpha  \exp \left[ - \frac{1}{2} \beta
  e_\alpha^2 \left(\fd-\vb)(\vr;\vr \right) \right]
\end{equation}

The second effect of step-$2$ resummations is a renormalization of the $f^*$ bonds in the final
diagrams $\cpp$. The summation operates for each pair of points in
$\cpp$ independently. (See for instance \cite{ACP94} for similar
topological considerations.) The sum of all possible single chains made
of $\fcc$ bonds and whose intermediate points carry weights $\zb^{[1]}$
is 
\be
\label{defFcc}
\Fcc(\cp;\cp') = - \beta e_\alpha e_{\alpha'} \fd ( \vr;\vr')
\end{equation}
The sum of the bond $(1/2) \left[ \fcc \right]^2 (\cp;\cp')$ and of all
subdiagrams made of the product of a Coulomb chain (possibly made of
only one bond $\fcc(\cp;\cp')$) and a Coulomb
chain with at least one intermediate point linking $\cp$ to $\cp'$ is
merely
\be
\frac{1}{2} \left[ \Fcc \right]^2 (\cp;\cp')
\end{equation}
The sum of the bond $\ft (\cp;\cp')$, its product by at least one
Coulomb chain with at least one intermediate point, and the product of
at least three Coulomb chains, one of which at least contains one
intermediate point leads to the simple resummed bond (see
Ref. \cite{Meer58} and \cite{Cornu96I} )
\be
\label{defFrt}
\Frt \equiv e^{-\beta \vsr + \Fcc} - 1 -\Fcc-\frac{1}{2} \left[ \Fcc
\right] ^2 
\end{equation}

After resummations the diagrams $\cpp$ are made of bonds $\Fcc$, 
$(1/2) \left[ \Fcc \right]^2$, and $\Frt$ with at most one bond between two points
(see Fig.\ref{f4}) and the following rules. A point -- different from the root point -- 
which carries only two bonds $\Fcc$ has a weight $\zbad - \zbau$ ($\rr_1$) 
(see Fig.\ref{f5}).  A point -- different from the root point -- which is linked only to a
$(1/2) \left[ \Fcc \right]^2$ bond has also a weight  $\zbad - \zbau$
($\rr_2$) (see Fig.\ref{f4}). The rules $\rr_1$ and $\rr_2$ arise from the definition 
of the Coulomb points which disappear in the resummation process and from rule 
$\rr$ of the first step of the resummations. All other points have a
weight $\zbad$. For instance, the diagrammatic representation of 
$\roa (\vr)$ in terms of $\cpp$ diagrams starts as
\begin{multline}
\label{diagrhoresum}
\roa(\vr)= \zbad (x) \exp \left\{ \sum_{\gamma} \int d\vr' \zbgd(x')
  \left[ \Fcc + \Frt \right] (\cp;\cp') + \right. \\
\left.  \sum_{\gamma} \int d\vr'
  \left[ \zbgd(x') - \zbgu(x') \right] \frac{1}{2} \left[ \Fcc \left( 
\cp;\cp' \right)  \right]^2 + \cdots \right\}
\end{multline}
The diagrams represented by the  dots contain more than one internal point.


\section{Screened potentials}
\label{section4}

\subsection{Partial derivative equations}

The screened potential $\fd$, as well as the other auxiliary object
$\fu$, are defined through integral equations \eqref{intrelphi1}. Since $\vw$ is a solution
of Poisson equation \eqref{pois}, \eqref{intrelphi1} is equivalent to a set of
local equations. For $x'>0$, the latter ones read
\begin{equation}
\left\{ \begin{array}{ll}
\Delta_\vr \fu (\vr;\vr') - \kudb (x) \fu (\vr;\vr') = - 4 \pi
\delta(\vr-\vr') & \mbox{if $x>0$} \\
\Delta_\vr \fu (\vr;\vr') = 0 & \mbox{if $x<0$}\rule{0mm}{6mm}
\end{array}
\right.
\label{eqphi1}
\end{equation}
with
\begin{equation}
\kudb (x) \equiv 4 \pi \beta \sum_\alpha \ead \zb_\alpha (x)
\label{defk1}
\end{equation}
According to its definition \eqref{defphi1}, $\fu$ is real. Since the operator $\left[ \Delta_\vr - \kudb (x) \right]$ is
self-adjoint, the real function $\fu(\vr;\vr')$ is symmetric under
exchange of its arguments when $\vr$ and $\vr'$ are in the same region,
\begin{equation}
\label{phiun}
\fu (\vr;\vrp) = \fu (\vrp;\vr) 
\end{equation}
Moreover, the invariance of the system under translations in directions
parallel to the plane interface implies that 
\be 
\fu (\vr;\vrp) = \fu (x,x',\vy-\vyp)
\end{equation}
where $\vy$ is the projection of $\vr$ onto the plane perpendicular to
the $x$-axis. 
According to its definition \eqref{defphi1}, $\fu$  obeys the
same boundary conditions as the electrostatic potential $\vw$: $\fu$ is
continuous in all space,
\be 
\lim_{x \rightarrow 0^-} \ew \frac{\partial \fu}{\partial x}
(x,x',\vy-\vyp) = \lim_{x \rightarrow 0^+} \frac{\partial \fu}{\partial
  x} (x,x',\vy-\vyp)
\label{boundcond}
\end{equation}
Since $\kudb (x)$ contains a Heaviside distribution at $x=b$ but no Dirac
distribution, the derivative of $\fu (x,x';\vy-\vyp)$ is continuous at
$x=b$. Moreover $\fu (x,x';\vy-\vyp)$ tends to zero when $x$ goes to
$+\infty$ or $-\infty$.
All previous properties are also true for $\fd (\vr;\vrp)$ with the only
difference
\begin{equation}
\kddb (x) \equiv 4 \pi \beta \sum_\alpha \ead \zbau (x) 
\label{defk2}
\end{equation}


\subsection{Scaling property}
\label{generalsol}

A Fourier transform allows one to change the system of partial derivative
equations \eqref{eqphi1} for each $\phi_j (\vr;\vrp)$ (with $j=1,2$)
into a system of one-dimensional differential equations with respect to
$x$. For $x'>0$,\\ 
\begin{subequations}
\label{eqphibis}
\begin{align}
 \left[ \rule{0mm}{6mm} \right.&\left. \frac{\partial^2}{\partial \xt^2} -
  \left(1+\vq^2\right)-U_{\rm j} 
  (\xt) \right] \fjt (\xt,\xt',  \vq)
    =
   -4 \pi \delta \left( \xt -
  \xt' \right) 
  & \mbox{ if $b<x$} \label{eqphibisa}\\
  \left[ \rule{0mm}{6mm} \right.&\left. \frac{\partial^2}{\partial
  \xt^2} -  \vq^2 \right] 
  \fjt (\xt,\xt',  \vq) 
   = -4 \pi \delta \left( \xt -
  \xt' \right) 
   & \mbox{ if $0<x<b$} \label{eqphibisb}\\
 \left[ \rule{0mm}{6mm} \right.& \left.\frac{\partial^2}{\partial \xt^2}
  - \vq^2 \right] 
  \fjt (\xt,\xt',  \vq) 
   = 0 
   & \mbox{if $x<0$} \label{eqphibisc}
 \end{align}
\end{subequations}
In \eqref{eqphibis} we have introduced the dimensionless variable $\xt =
\kzj x$ with 
\begin{eqnarray}
\kzu & \equiv & \sqrt{ 4 
 \pi \beta \sum_\alpha \ead z_\alpha } \label{defku}\\
\kzd & \equiv & \sqrt{ 4 
 \pi \beta \sum_\alpha \ead z_\alpha
  e^{\epsa} } \label{defkzd}\\
  \epsa & \equiv & \frac{1}{2} \beta \ead \kzu\label{defepsilonalpha}
\end{eqnarray}
and the dimensionless Fourier transform 
\begin{equation}
\label{scalephi}
\tilde{\phi_{{\rm j}}} (\xt,\xt',\vq) =
\kzj \phi_{{\rm j}} (x,x',\kzj \vq)
\end{equation}
 with
$\phi_{{\rm j}} (x,x',\kzj \vq ) \equiv \int d\vy \exp
\left[ i \kzj \vq.\vy \right] \phi_{{\rm j}}(x,x',\vy)$.
According to \eqref{ztheta} and \eqref{defk1}, for $\xt = \kzu x$
\be
U_1 (\xt) = \frac{4 \pi \beta}{\kzu^2} \sum_\alpha \ead z_\alpha
\left(  e^{\dw \epsa/ (2 \xt )} - 1 \right)
\label{defU1}
\end{equation}
while, according to \eqref{valuez1} and \eqref{defk2}, for $\xt = \kzd x$
\be
U_2 (\xt) = \frac{4 \pi \beta}{\kzd^2} \sum_\alpha \ead z_\alpha
\left[e^{-(1/2) \beta \ead \left( \fu - \vb \right) (\vr;\vr)} -
e^{\epsa}\right]
\label{defU2}
\end{equation}
When $x$ goes to $\infty$, $\fu$ tends to the bulk value $\fub$ which is the
solution of 
\be
\label{eqphibulk}
\left[\Delta_{\vr} - {\kud}\right] \fub(\vr,\vr')=
-4\pi\delta(\vr-\vr')
\end{equation}
for all $\vr$'s. As rederived below,
\be
\fub(\vr,\vr')=\kzu \tilde{\fbz}(\kzu | \vr - \vr' | )
\end{equation}
with
\be
\label{valuephibulkr}
\tilde{\fbz}(\tilde{\vr},\tilde{\vr'})
=\frac{e^{- \vert\tilde{\vr}-\tilde{\vr'}\vert}}
{\vert\tilde{\vr}-\tilde{\vr'}\vert}
\end{equation}
As a consequence,
\be
\left( \fub - \vb \right) (\vr;\vr) = - \kzu
\end{equation}
and the definitions \eqref{defU1} and \eqref{defU2} ensure
that
\be
\lim_{\xt \rightarrow + \infty} U_{\rm j} (\xt) = 0 
\end{equation}
The boundary conditions for $\tilde{\phi}_{\rm j} \xxq$ along the $x$-axis
are the same as for $\phi_{\rm j}(x,x',\vy-\vy')$ (see
\eqref{boundcond}).
Moreover, since $\kzd \equiv \kzd ( \kzu, \eps)$ a mere scaling analysis
of \eqref{eqphibis} shows that 
\begin{equation}
  \label{}
  \phi_{j} (\vr, \vr' ; \kzj, \beta e^2, b, \dw ) = \kzj  \tilde{\phi_j} 
 (\kzj \vr, \kzj \vr ' ; \eps, \kzj b , \dw)
\end{equation}
where $e^2$ ($\eps \equiv (1/2) \kzu \beta e^2$) is the generic notation
for the $\ead$'s ($\epsa$'s).


\subsection{General solutions}
\label{sec4.3}

In the following only the value of $\fj(x,x';\vy-\vy')$ for $x'>b$ will be
involved and \eqref{eqphibisb} becomes an homogeneous equation as well as
\eqref{eqphibisc}. The simple general solutions of the latter equations
that satisfy boundary conditions take the form
\begin{equation}
\label{phijtild}
\fjt (\xt,\xt',\vq; \bt )= \left\{ 
\begin{array}{ll}
B_{{\rm j}} (\xt';q;\bt) \left( 1 - \dw \right) e^{q \xt} 
& \mbox{ if $ x <0$}\\
B_{{\rm j}} (\xt';q;\bt) \left[ e^{q \xt} + \dw e^{-q \xt} \right]
& \mbox{ if $ 0< x <b$}
\end{array} 
\right.
\end{equation}
where $q\equiv \vert\vq\vert$. The general solution of \eqref{eqphibisa} 
is the sum of a particular solution $\fjt^* (\xt,\xt',\vq)$
and the general solution $h_j $  of the corresponding homogeneous
equation
\begin{equation}
\frac{d^2 h_{{\rm j}}}{d \xt^2}(\xt;\vq) + \left[ \eq - \uj (\xt)
\right] h_{{\rm j}}(\xt;\vq) = 0 
\label{eqhomo}
\end{equation}
with $\eq \equiv - (1 + \vq^2)$. We look for a solution $\hjps$ ($\hjms$) 
which vanishes (blows up)  at large positive $x$,
\begin{equation}
  \label{limhp}
  \lim_{x \rightarrow + \infty} h_{{\rm j}}^{+} (\xt; \vq) = 0 
\end{equation}
while $ \lim_{x \rightarrow + \infty} h_{{\rm j}}^{-} (\xt; \vq) = +\infty $.

Since the $\uj (\xt)$'s vanish when $\xt$ goes to $+\infty$, 
then, for any given positive number $ \eta >0$
there exists some $x_{{\rm q}, \eta}$ such that for all
$x>x_{{\rm q},\eta }$ : $\eq - \uj (\xt)  \leq - |\eq | + \eta$. 
As a consequence
(see Ref.\cite{Messiah1}) there exists one particular solution $\hjp (\xt;\vq)$
(with a particular multiplicative constant) which tends to zero when $x$
goes to $+\infty$ at least as fast as $\exp \left[ - \xt \race 
\right] $,
\begin{equation}
\forall \xt \geq \xt_{{\rm q}, \eta} \quad |\hjp (\xt;\vq)| \leq 
 \exp \left[ - \xt  \race \right]
\label{ineg}
\end{equation}
whereas all other solutions grow to infinity at least as fast as
\linebreak
  $\exp \left[ \xt \race \right] $. 
Let call $\hjm (\xt;\vq)$ such a particular solution which diverges when
$\xt$ goes to $+\infty$.
Henceforth (see for instance 
\cite{Zwillinger}) a particular solution of \eqref{eqphibisa} 
for $x>0$ is just 
\begin{equation}
\label{cdsolpart}
\fjtp (\xt,\xt',\vq) \equiv - \frac{4 \pi}{W_{{\rm j,q}}} \, \hjm
  \left(\inf(\xt,\xt');\vq\right) \hjp \left(\sup(\xt,\xt');\vq\right)
\end{equation}
where the Wronskien $W_{{\rm j, q}} \equiv \hjm (\xt) (d\hjp(\xt)/d\xt)
-\hjp (\xt) (d\hjm(\xt)/d\xt) $ is independent from $\xt$ because $
d^2/d\xt^2 + \left[ \eq - \uj(\xt) \right]$ is a self-adjoint
operator.

Finally, a generic solution of \eqref{eqphibisa} which vanishes when $x$
goes to $+\infty$ takes the form 
\begin{equation}
  \label{}
  \fjt (\xt,\xt',\vq ;\bjt) = \fjtp (\xt,\xt',\vq) + A(\xt';\vq, \bjt)
  \hjp (\xt;\vq)
\quad \mbox{ for $b < x$ }
\end{equation}
with $\bjt \equiv \kzj b$.
Moreover the symmetry property \eqref{phiun} implies that 
the solution of \eqref{eqphibisa} with
adequate boundary conditions at $+\infty$ may be written as 
\begin{equation}
\label{decomphi}
\fjt(\xt,\xt', \vq;\bjt) = 
\fjtp (\xt,\xt',\vq) + Z(\vq, \bjt) \hjp (\xt;\vq) \hjp (\xt';\vq)
\end{equation}
for $b<x$ and $b<x'$. The coefficient $Z(\vq, \bt)$ is
determined by the continuity relations at $\xt = \bjt$ and $\xt = 0$
with the solutions \eqref{phijtild} for $0<x<b$ and $x<0$.


\subsection{Exact solutions for $\fu$ in two special cases } 
\label{sec6.1}

An exact analytic expression -- derived from the general method  in Section
\ref{sec4.3} --
exists for the screened potential $\fu$ in two cases. The first one
corresponds to $\dw=0$ so that  
the wall has no electrostatic response for any strength $\eps$ of the
Coulomb coupling inside the fluid. The second situation is the strict
limit $\eps = 0$ for $\dw \neq 0$. 
Indeed, in both situations
$U_1 = 0$  and when $x'>b$ 
$\fut $ is the solution $\ftz (\xt,\xt',\vq;\bt,\dw)$ of
\begin{subequations}
\label{eqphi1pw}
 \begin{align}
 \left[ \rule{0mm}{6mm} \right.&\left. \frac{\partial^2}{\partial \xt^2} -
  \left(1+\vq^2\right)  \right] \ftz (\xt,\xt', \vq ;\bt, \dw)
    =
   -4 \pi \delta \left( \xt -
  \xt' \right) 
  & \mbox{ if $b<x$} \label{eqphi1pwa}\\
  \left[ \rule{0mm}{6mm} \right.&\left. \frac{\partial^2}{\partial
  \xt^2} -  \vq^2 \right] 
  \ftz (\xt,\xt', \vq;\bt, \dw) 
   = 0
   & \mbox{ if $x<b$} \label{eqphi1pwb}
 \end{align}
\end{subequations}
In the following we use the notation $\bt$ for $\kzu b$ and introduce
$\bjt$ only when we consider both $\kzu b$ and $\kzd b$.

When $\dw=0$, for any coupling constant $\eps$
\begin{equation}
  \label{}
  \widetilde{\phi}_1 (\xt,\xt', \vq; \eps, \bt, \dw = 0) = \ftz
  (\xt, \xt', \vq; \bt, \dw =0 )
\end{equation}
and $\fut$ has no special property at $x=0$. (According to
\eqref{boundcond}, $\fut$ and $\partial \fut / \partial \xt$ are
continuous at the crossing of the wall.) In fact, up to a translation
of its argument $x$ equal to $b$, $\fut$ is defined as the
mean-field potential $\phi$ in the linearized Poisson-Boltzmann approximation 
in the case of a
multicomponent plasma in the vicinity of a plain hard wall ($\dw =0$)
located at $x=0$. Such a wall exerts only a geometric constraint without
any electrostatic attraction : there is no need for introducing the
hard-core repulsion $\Vsr$ \eqref{defVsr} involved in our generic model, so
that $b$ could be set to zero in this particular case. 
(We recall that at leading order in $\eps$, the correlation (or Ursell) function   of charges with a hard-core diameter $\sigma$ is just
equal to $-\beta e_\alpha e_{\alpha'} \phi (x,x';\vy - \vy ';\dw=0)$
\cite{Guer70} and
does not involve $\sigma$.) 
As recalled in \eqref{decomphi}
$\fu$ may be written as the sum
\begin{equation}
  \begin{split}
  \label{valuephi1pw}
  \ftz ( x,x',\kzu \vq;\bt , \dw = 0) = 
    &  \ftb (\kzu |x-x'|, \vq) \\
    & + \hthw (\xt + \xt ' - 2 \bt ;\kzu \vq) 
\end{split}
\end{equation}
(see  Ref. \cite{Janco82I,Russes}) where $\ftb$ is a particular 
solution of
\eqref{eqphi1pw} which is chosen to be its bulk value,
\begin{equation}
\ftb (|\xt- \xt '|,\vq) = \frac{2 \pi}{\rac}
\, e^{- |\xt - \xt '| \rac}   
\label{valuephib0}  
\end{equation}
and $\tilde{h}_{{\scriptscriptstyle H \, W}}^{+ }$ is a solution of the 
associated homogeneous equation which vanishes
when $x$ goes to $+\infty$ and whose
coefficient is entirely determined by the boundary conditions at the
interface,
\begin{equation}
  \label{valuehop}
 \hthw (\xt + \xt' - 2\bt ;\vq) =  
 \frac{2 \pi}{\rac} \, \frac{\rac -
  |\vq|}{\rac + |\vq|} e^{- (\xt + \xt '-2 \bt) \rac}
\end{equation}
However $\phi_2 ( \dw = 0)$ cannot be calculated explicitly, because
$U_2(\xt) \neq 0$
according to \eqref{defU2} and \eqref{valuephi1pw}.
 
When $\dw \neq 0$, the electrostatic response of the wall disappears in
the limit $\eps = 0$ for $\bt$ fixed,
\begin{equation}
  \label{}
  \lim_{\eps \rightarrow 0} \widetilde{\phi}_1 (\xt,\xt',\vq; \eps,
  \bt, \dw ) = \ftz (\xt,\xt',\vq; \bt, \dw)
\end{equation}
where
\begin{multline}
 \label{defphi10}
  \ftz(\xt,\xt ',\vq;\bt , \dw) =  \ftb (|\xt - \xt '|, \vq) \\
          + \Zv \hthw (\xt + \xt '- 2\bt ; \vq)
\end{multline}
$Z$ is a renormalization factor arising from the continuity conditions at
$x=b$ when $\dw \neq 0$, 
\begin{equation}
  \label{defZ}
  \Zv \equiv \frac{ 1 - \dw e^{-2 q \bt } \left[ \rac + \vert\vq\vert
  \right]^2}{1 - \dw e^{-2 q \bt } \left[ \rac - \vert\vq\vert
  \right]^2} 
\end{equation}


\subsection{General equivalent integral equations}
\label{cdmethodgen}

According to section \ref{sec4.3} the determination of a screened
potential $\phi$ is equivalent to
solving the homogeneous equation \eqref{eqhomo}.
The asymptotic equation at large distances associated with \eqref{eqhomo}
has two exact linearly independent solutions
$A_{\pm} \exp \left[ \mp \xt  \rac  \right]$ which either 
vanishes or diverges when $\xt$ tends to $+ \infty$. 
Thus, in the following $h_{{\rm j}}^{\pm \, *}$ will be looked for 
under the form 
\begin{equation}
h_{{\rm j}}^{\pm \, *} (\xt;q,\bt) = e^{\mp \xt \rac} \left[ 1 + H_{{\rm
      j}}^{\pm \, *}
  \left( \xt ; \rac  , \bt \right) \right]
\label{defH}
\end{equation}
$h_{{\rm j}}^{\pm \, *}$ is defined up to a multiplicative constant.
For convenience sake, we choose the particular solution such that
$ h_{{\rm j}}^{\pm \, *} ( \bt; q, \bt ) = \exp \left( \mp \bt \rac
\right)$.
 
When $\uj (\xt)$ vanishes at large distances at least as
fast as $1/\xt$, we show in Appendix \ref{AppendiceA} that $\Hjpe (\xt)$ is
the only solution of the integral equation 
\begin{equation}
  \label{intrelHp}
  \Hjpe (\xt) = - \LUj \left[ 1 + \Hjpe ;2 \rac , \bt   \right] (\xt)
\end{equation}
where $\LUj$ is 
a linear operator operating on a function $f$ as 
\begin{equation}
  \label{defop}
  \LUj \left[ f; \gamma, \bt \right] (\xt) \equiv \int_{\bt}^{\xt} dv \,
  e^{\gamma v} \, \int_v^{+\infty} dt \, e^{- \gamma t} U_{\rm j} (t)
  f(t) 
\end{equation}
As it is the case for Dyson
equation, the solution $\Hjpe$ of \eqref{intrelHp}, denoted by $\HUj$ in
the following, can be written as the formal series
\begin{equation}
  \label{series}
  \Hjpe = \HUj \equiv - \LUj [1] + \LUj \left[ \LUj [1] \right] - \LUj
  \left[ \LUj \left[ \LUj [1] \right] \right] + \cdots
\end{equation}
 For the sake of conciseness we
have omitted the dependence upon the parameters $\gamma = 2 \rac$ and
$\bt$ in \eqref{series}. 

We notice that, similarly, 
the function $\hjm$ which explodes exponentially fast when
$x$ tends to $+\infty$ and which is equal to $\exp [ \bt \rac ] $
at $\xt=\bt$ corresponds to a $\Hjme (u;q)$ which obeys the integral equation
\begin{multline}
\Hjme (u;\rac, \bt) = \int_{\bt}^u \, dv \, 
 e^{-2 v \rac } \int_{\bt}^{v} dt \, e^{2t \rac }
  \, \uj (t) \\
 \times \left[ 1+\Hjme (t;\rac ,\bt) \right] 
  \label{intrelHm}  
\end{multline}
A series similar to \eqref{series} can be written for $\Hjme$.


\subsection{Structure of the $\eps$-expansions of screened potentials}
\label{sectionquatresix}

The structure of the $\eps$-expansions of $\fu$ and $\fd$ can be
investigated by means of the series representation \eqref{series} for
the intermediate object $\HUj$ combined with bounds upon $\Uj$. First we
consider $U_1 (\xt)$ defined in \eqref{defU1}, 
\begin{equation}
  \label{}
  U_1 (\xt) = \sum_\alpha \uua 
  \left[ \exp\left(\frac{\dw \epsa}{2 \xt}\right) - 1 \right]
\end{equation}
with $\uua \equiv 4 \pi \beta \ead z_\alpha / \kud$.
As shown in subsection \ref{cdappenu}
\begin{eqnarray}
  \label{}
|U_1 (\xt) | \leq  \frac{|\dw|}{2}  \left( \sum_\alpha \uua \epsa \right)
 \frac{1}{\xt} \quad \rm{ if } \, \dw < 0  \\
|U_1 (\xt) | \leq  \frac{\dw}{2}  \left( \sum_\alpha \uua \epsa 
\exp\left(\frac{\dw
 \epsa}{2 \bt}\right) \right)
 \frac{1}{\xt} \quad \rm{ if } \,\dw > 0 
\end{eqnarray}

The result of the $\eps$-expansion for 
the solutions $h_{{1}}^{\pm \, *}$ of the homogeneous equation \eqref{eqhomo} 
proves to be very simple at first order in $\eps$, as shown in Appendix B. 
The term of order $\eps$ in the $\eps$-expansion of $h_{1}^{\pm \, *} $ coincides with the 
corresponding term in the $\eps$-expansion of the solution of the equation \eqref{eqphibisa}
where $U_1$ is replaced by its linearized value $\Uulin$
    \begin{equation}
       \label{cddefuulin}
       \Uulin (\xt;\eps) \equiv \left( \sum_\alpha \uua \epsa \right)
      \frac{\dw}{2 \xt}
    \end{equation}
(We stress that such a coincidence is no longer valid at higher orders in
$\eps$.) More precisely~:
\begin{multline}
  \label{valuehfin}
  \hupe (\xt; q,\bt) = e^{-\xt \rac} \left[ 1 + \rule{0mm}{5mm} \right. \\ 
 \left. \int_{\bt}^{\xt} dv \,
  e^{2 v \rac} \int_v^{+\infty} dt \, e^{-2 t \rac} \Uulin (t;\eps) \right]
   + \Oexp(\eps^2)
\end{multline}
where $\eps$ is a symbolic notation for the dependence on the
$\eps_\alpha$'s.
The double integral involving $\Uulin (t;\eps)$ in \eqref{valuehfin} is
proportional to $\eps$, while $\Oexp(\eps^2)$ is equal to $\eps^2$ 
-- possibly multiplied by some power of $\ln \eps$ -- times 
a function $f(\xt; \eps, \bt)$ which decays
exponentially fast over the length-scale $\rac$ and which is bounded for
all $x>b$ by a function of $\bt$ and $\eps /\bt$. 
The function $f(\xt; \eps,\bt)$ may contain powers  $\xt^{p(\eps)}$
(with $\lim_{\eps \rightarrow 0} p(\eps) = 0$)
times an exponential of $ -
\xt \rac$, as it is the case for the exact solution of the equation
\eqref{eqhomo} where $U_1 (\xt)$ is replaced by $\Uulin (\xt)$ (see
\eqref{Has}).

We point out that the correction of
order $\eps$ in $\hjep (\xt; q, \bt)$ is a function of $\xt$ only,
whereas the higher-order corrections are functions which vary on both
scales $1$ and $\eps$ :
\begin{equation}
 \label{dlepsd}
  \hupe (\xt) = h_{{1}}^{(0) \, +} (\xt) + \eps h_{{1}}^{(1) \,
  +}(\xt) +\eps^2 h_{{1}}^{(2) \, +} \left(\xt, \frac{\xt}{\eps} \right) 
+ {\cal O} (\eps^3) 
\end{equation}
This property ensures that the zeroth order term in the
$\eps$-expansion of the derivative $\partial \hupe / \partial \xt$
depends only on the corresponding term in the $\eps$-expansion of
$\hupe$, whereas the first-order term involves the derivatives of both
the first and second corrections to $\hupe$,
\begin{equation}
  \label{derivative}
  \frac{\partial \hupe}{\partial \xt} = \frac{d h_{{1}}^{(0) \, +}
  (\xt) }{d \xt} + \eps \left [ \frac{d h_{{1}}^{(1) \, +}
  (\xt) }{d \xt} + \left. \frac{\partial h_{{1}}^{(2) \, +}
  (\xt,u) }{\partial u} \right|_{u = \xt / \eps} \right] 
 + {\cal O} (\eps^2)
\end{equation}
According to \eqref{intrelHm}, results similar to \eqref{valuehfin} and
\eqref{dlepsd} also hold for $\hume (\xt)$.

Finally, we turn to the $\eps$-expansion for the screened potential $\fu$.
 $\fu$ is calculated from the $\hupme$'s by using the expression
 \eqref{decomphi} valid for $x>b$ together with
the continuity relations \eqref{boundcond} involving the solutions
  \eqref{phijtild}  in
  the region $x<b$. As a consequence of \eqref{derivative}, at leading
  order in $\eps$, the
  continuity relation \eqref{boundcond} for the derivative $\partial \fu /
  \partial \xt$ involves only the $h_{{1}}^{(0) \pm}$'s and their
  derivatives $d h_{{1}}^{(0) \pm}/ d \xt$ .
  Therefore, the leading term $\tilde{\phi}_{{1}}^{(0)}$ 
in the $\eps$-expansion of $\fu$ is entirely determined by
  the functions $h_{{1}}^{(0) \, +}$ and $h_{{1}}^{(0) \, -}$
  (and not by functions which appear at higher orders in the
  $\eps$-expansions of the $h_{{1}}^{* \, \pm}$'s) and it coincides with
 $\ftz$ given in \eqref{defphi10}, 
\begin{equation}
  \label{analysphiu}
  \fu (\vr,\vr') = \kzu
  \tilde{\phi}^{(0)} (\kzu \vr , \kzu \vr';
   \kzu b, \dw) 
  + \kzu \Oexp(\eps)  
\end{equation} 
where $\Oexp(\eps)  $ has the same meaning as in \eqref{valuehfin}. As a consequence of \eqref{analysphiu}
\begin{equation}
  \label{propself}
  - \frac{1}{2} \beta \ead \left[ \fu - \fbz \right] (\vr, \vr) = \epsa
    F(\xt; \eps, \bt)
\end{equation}
where $F(\xt; \eps, \bt)$ is a continuous function of $\xt$ in the
interval $ b \leq \xt < \infty$ which 
decays exponentially fast (with possible multiplicative powers
$x^{p(\eps)}$) over a typical length $1$ when $\xt$ goes to
$+\infty$. $F(\xt; \eps, \bt)$ is integrable when $\xt$ goes to
$+\infty$ even when $\eps$ vanishes, and may be bounded as follows,
\begin{equation}
  \label{}
  \left| F (\xt; \eps, \bt) \right|  \leq M_F (\eps / \bt) g (\xt) \qquad \forall
  \xt \geq \bt
\end{equation}
where 
$g(\xt)$ is  continuous and integrable in the interval $\left[ \bt, +
  \infty \right[ $.

>From the previous result, we get the structure of $U_2 (\xt)$ given in
\eqref{defU2}. For $\xt = \kzd x$
\begin{equation}
  \label{}
  U_2 (\xt) = \sum_\alpha \uda \left[ e^{\epsa F (\xt (\kzu/\kzd); \eps,
  \bt)} - 1 \right]
\end{equation}
with $\uda \equiv 4 \pi \beta \ead z_\alpha \exp (\epsa) /\kzd^2$. We
show in Appendix \ref{AppendiceB} that, as it is the case for $U_1$, the first
terms in the $\eps$-expansion of $\hdpe$ are given by \eqref{valuehfin}
where $\Uulin (\xt;\eps)$ is replaced by 
\begin{equation}
  \label{Udlin}
  \Udlin (\xt; \eps, \bt) = \sum_\alpha \uda \, \epsa \, F(\xt \kzu/\kzd; \eps,
  \bt ) 
\end{equation}
The double integral involving $\Udlin (t; \eps, \bt)$  in \eqref{valuehfin}
is not merely
proportional to $\eps$. According to definitions \eqref{defku} 
and \eqref{defkzd}
\begin{equation}
  \kzd = \kzu \left[ 1 + \cO (\eps) \right]
\end{equation}
and the properties of the double integral at stake imply that the first
two terms in the $\eps$-expansion of $\hdpe$ are given by
\eqref{valuehfin} where $\Udlin$ is replaced by
\begin{equation}
  \label{}
  \left( \sum_\alpha \uda \, \epsa \right)  \, F(\xt; \eps=0 ,  \bt)
\end{equation}
Subsequently, all properties derived from \eqref{valuehfin} also hold
for $\hdpe$.

Eventually, 
the main result of the previous perturbative analysis is that, at leading 
order in $\eps$, $\fu$ and
$\fd$ coincide with the same function $\phi^{(0)}$ apart from a scaling
dependence upon either $\kzu$ or $\kzd$
\begin{equation}
  \label{analysphi}
  \fj (x,x',\kzj \vq; \eps, \kzj b, \dw) = \frac{1}{\kzj}
  \tilde{\phi}^{(0)} (\kzj x, \kzj x',\vq;\kzj b, \dw) 
 + \frac{1}{\kzj} \Oexp(\eps)  
\end{equation}
In \eqref{analysphi} $\Oexp ( \eps )$ denotes a function of the variable
$x$ which satisfies two properties. First it decays exponentially fast
over the scale $\kzj$ in the sense that it falls off as $\exp (-\kzj x
\rac)$ times a function which may increase as $x^{p(\eps)}$ with
$\lim_{\eps \rightarrow 0} p (\eps) = 0 $. Second, $\Oexp(\eps)$
remains bounded by $\eps$ -- possibly multiplied by some power of $\ln \eps$ --
times a function of $\eps/(\kzu b) \varpropto
\beta e^2 /b$ for all
$x>b$, even if $\Oexp(\eps)$ may drastically vary over the scale $\eps
\kzj \sim \beta e^2 \ll \kzj$.

As shown in next section, because of the neutrality constraint \eqref{zneut},
the explicit value of the correction of order $\eps$ in $\fd$ happens
not to appear in the first correction to the density profile. (Only some
properties of it must be known in order to settle that it really does
not contribute to the density profile at order $\eps$). The latter
result may be viewed as a consequence of the following property, already
used in Ref.\cite{Guer70}. According to the first equation of the 
BGY hierarchy,
the first-order correction induced by Coulomb interactions in the
density profile is determined only by the potential drop $\Phi(x)$ and 
the Ursell function at leading
order; besides the latter one is nothing but the screened potential $\fd$
in the limit $\eps = 0$. We recall that $\lim_{\eps \rightarrow  0} \fd$ 
is drastically
different from the bare long-ranged Coulomb potential $\vw$, which would
be a too crude approximation for the Ursell function~: though 
$\lim_{\eps \rightarrow 0} \fd $ has the same amplitude $\beta e^2$ as 
$\vw$, it is a short-ranged function with a characteristic scale 
$\kappa \varpropto \eps/(\beta e^2)$.


\section{Scaling analysis in the weak-coupling limit}
\label{section5}

In the weak-coupling regime of interest, the particle density $\roax$ 
at leading order in $\eps$ is equal to its value $\rho_\alpha^{{\rm id}} (x)$
in an ideal gas submitted to the external
potential corresponding to the self-energy $- \dw \ead / 4x$ of a charge
in the presence of a wall with an electrostatic response,
\begin{equation}
  \label{roid}
  \rho_\alpha^{{\rm id}} (x) = z_\alpha \exp \left( \dw \beta \ead / 4 x \right)
\end{equation}
According to \eqref{defepsilon} and \eqref{defku}, the small dimensionless
coupling parameter inside the Coulomb fluid may be chosen as
\begin{equation}
\label{defepsa}
 \eps_\alpha \equiv \frac{1}{2} \kzu \beta \ead \ll 1 
\end{equation}
while the coupling constant with the dielectric wall
\begin{equation}
  \label{}
  \frac{\dw \beta \ead}{4 b }
\end{equation}
can take any finite given value.

\subsection{Screened fugacities}

In the weak-coupling regime,
a simple scaling analysis may be performed in the bulk as in the inhomogeneous
situation near the wall. 
As a consequence of \eqref{analysphi}
 $(1/2) \beta e_{\alpha}^2(\fu-\vb)(\vr,\vr)$ scales as 
$\epsa$ at leading order and takes the generic form
\begin{equation}
  \label{valueself}
 -\frac{1}{2} \beta \ead \left( \fu - \vb \right) (\vr;\vr) =  \epsa 
 \left\{ 1  - L(\kzu (x-b);\kzu b, \dw) \right\} 
 + \Oexp \left( \eps^2 \right)   
\end{equation}
where the term of leading order is entirely determined by \hfill\break 
$\kzu \tilde{\phi}^{(0)} (\kzu \vr,  \kzu \vr;  \kzu b, \dw)$,
\begin{equation}
  \label{}
  L(\xt-\bt;\bt, \dw) = 
  \int \frac{d^2 \vq}{(2 \pi)^2} 
 \left[ \phit^{(0)}
   (\xt,\xt,\vq; \bt ,\dw) - \ftb(\xt,\xt,\vq) \right]
\end{equation}
$L(\xt-\bt;\bt, \dw)$ has been studied in Section 3.2 of Paper I. It can be decomposed
into
\begin{equation}
  \label{decompL}
  L(\xt - \bt; \bt, \dw) = - \dw \frac{e^{-2 \xt}}{2 \xt} + \Lb (\xt ; \bt
  , \dw)
\end{equation}
where $\Lb (\xt-\bt ; \bt, \dw)$ remains finite even when $\xt = \bt = 0$.
Thus, according to \eqref{valuez1}, \eqref{valueself} and
\eqref{decompL},  $\zbau$ proves to read
\begin{multline}
  \label{crucial}
\zbau(x)=z_\alpha \, e^{\epsa} \, \theta (x-b) 
\exp\left[\dw \frac{\beta e_{\alpha}^2}{4x}e^{-2\kzu x}\right]\\
\times
\left\{ 1-\epsa \Lb(\kzu x; \kzu b, \dw)+ \Oexp\left( \eps^2\right)\right\}
\end{multline}

The expression \eqref{valuez2} of $\zbad$ differs from
the expression \eqref{valuez1} of $\zbau$ only by the replacement of $\fu$ by
$\fd$.
At leading order in $\eps$, apart from the change of $\kzu$ into 
$\kzd$,
$\fd$ coincides with $\fu$ 
(see \eqref{analysphi}).  
Thus $\zbad$ is given by \eqref{crucial} where 
$\kzu$ is replaced
by $\kzd$ and $\epsa$ is multiplied by  $\kzd / \kzu$.
However, according to \eqref{defkzd}, 
\begin{equation}
  \label{rapkappas}
  \frac{\kzd}{\kzu} \epsa = \epsa \left[ 1 + \Oeps \right]
\end{equation}
 so that
\begin{multline}
  \label{crucialbis}
\zbad(x)= \theta (x-b) z_\alpha \exp\left[\dw 
 \frac{\beta e_{\alpha}^2}{4x}
 e^{-2\kzd x}\right]
\left\{ 1+ \epsa \left[ 1 - \Lb(\kzd x; \kzd b, \dw) \right] \right.\\
 \left. + \Oexp\left( \eps^2\right) \right\}
\end{multline}
We remind the reader that $\Oexp\left( \eps^2\right)$ denotes a function which tends
exponentially fast to a constant of order $\eps^2$ -- possibly multiplied by
some power of $\ln \eps$ -- when $x$ goes to
$+\infty$.


\subsection{Diagrams contributing at leading order in $\eps$}
\label{cdanalysdiag}

First, we notice that even in the case where $L$ is not bounded at the origin 
-- as it is the case
when $\ew>1$ -- an $\eps$-expansion can be performed for integrals of the form
$ \int d\vr' \, \zbgd (x') \, f(\vr, \vr')$. For that purpose, we introduce
\begin{equation}
  \label{defwzero}
  \wza \equiv \exp \left[ \dw \frac{\epsa}{2 \xt } e^{-2 \xt} \right] - 1
\end{equation}
and we use the following formulas.
  Let us consider a dimensionless 
function $f$ of the variable $x$  which involves  two length scales $l_1$ and
$l_2$. $f$ may be written as $f(x/l_2; l_1/l_2)$. We set $\eps\equiv l_1/l_2$.
When $b>l_2$, for all $x>b$ $x/l_2$ remains finite when $\eps$ vanishes and  
\begin{equation}
\label{basicformdeux}
\Ee \left[ \int_{b}^{+ \infty} dx \, f ( x/l_2; \eps)
  \right] = l_2 \int_{b/l_2}^{+\infty} d\xt_2 \quad \Ee \left[ f (\xt_2;\eps)
  \right]
\end{equation}
When $b\ll l_2$, we 
use the identity
\begin{equation}
  \label{cdfracint}
  \underset{l_1 \ll l_2}{{\rm Exp}} \int_b^{\infty} dx \, \ldots = 
 \underset{l_1 \ll l \ll l_2}{{\rm Exp}} \left[ \int_b^l dx \, \ldots + 
 \int_l^\infty dx \, \ldots \right]
\end{equation}
When $l/l_2$ vanishes, $x/l_2$ goes to zero for all $x<l$ but remains
finite for any $x>l$. Thus
a basic formula for $\eps$-expansions when $b\ll l_2$ reads
\begin{equation}
\label{basicform}
\begin{split}
  \Ee \left[ \int_{b}^{+ \infty} dx \,  f ( x/l_2; \eps)\right] 
  =  \underset{l/l_1 \rightarrow + \infty}{\rm Exp}\,
  \underset{l/l_2 \rightarrow 0}{\rm Exp}  
  & \left\{ \rule{0mm}{6mm}
  l_1 \int_{b/l_1}^{l/l_1} d\xt_1 \quad \Ee \left[ f (\eps\xt_1;\eps)
  \right] \right. \\ 
 & \left.+  
 l_2 \int_{l/l_2}^{+\infty} d\xt_2 \quad \Ee \left[ f (\xt_2;\eps)
  \right] \rule{0mm}{6mm}\right\}
\end{split}
\end{equation}
We notice that only the sum of the two integrals in \eqref{basicform} is
independent from $l$, whereas the first (second) integral may diverge when $l/l_1$
becomes infinite (when $l/l_2$ vanishes). Whatever the value of $b$ may be,
\eqref{basicformdeux} and \eqref{basicform}   imply that 
$\kd \int d\vr' \, w_0(x'; \epsa,\dw) \, f(\vr, \vr')$ is at least of order 
$\eps \ln \eps$
(see also subsection 3.3 of Paper I).

The simplest diagrams contributing to $\roa (x)$ are written in
\eqref{diagrhoresum}. The potential involved in $\Fcc$ is $\fd$. First, the
scaling property of $\fd$ at leading order in $\eps$, namely
\begin{equation}
  \label{scale}
  \fd^{(0)} (\vr,\vr';\kzd) = \kzd \phit^{(0)} (\kzd \vr,\kzd \vr'),
\end{equation}
implies that, according to  \eqref{defFcc}, \eqref{defkzd}, \eqref{defepsa}
and \eqref{crucialbis},
\begin{equation}
  \label{sw1}
  \int d\vr' \, \zbgd (x') \, \Fcc (\vr;\vr') = \cO (\eps^0)
\end{equation}
Moreover, this leading correction depends on the species only through a 
coefficient $z_\gamma$. Subsequently, according to the neutrality constraint
\eqref{zneut}, after summation over species, the diagram with one bond
$\Fcc$  contributes only at next order in $\eps$,
\begin{equation}
  \label{toto}
  \sum_{\gamma}\int d\vr' \, \zbgd (x') \, \Fcc (\vr;\vr') = \cO (\eps)
\end{equation}

In the same way, the scaling argument shows that
\begin{equation}
  \label{tutu}
  \int \, d\vr' \,  \zbad(x')  \left[ \Fcc
  (\vr;\vr') \right]^2 = \cO (\eps) 
\end{equation}
An $\eps$-expansion can be performed for the contribution from $[\Fcc]^2$
 to the density representation \eqref{diagrhoresum}.
Since  $\Lb(\kzj x)$ decays exponentially fast, comparison of 
\eqref{crucial} and \eqref{crucialbis} implies with \eqref{rapkappas} that
\begin{equation}
  \label{difz}
\zbad (x) = \zbau (x)
\times \exp\left[\dw \frac{\beta e_{\alpha}^2}{2x}(e^{-2\kzd x}-e^{-2\kzu
x})\right]
+ \cO \left( z_\alpha \eps^2 \right)
\end{equation}
and a straightforward calculation leads to the result
\begin{equation}
  \label{}
  \int \, d\vr' \, \left[ \zbad(x') -\zbau(x') \right] \left[ \Fcc
  (\vr;\vr') \right]^2 = \cO (\eps^2) 
\end{equation}

The basic formulas \eqref{basicformdeux} and \eqref{basicform}  with $l_1=\beta e_\alpha e_\gamma$ and 
$l_2= \kzd^{-1}$ allow one to show that
\begin{equation}
  \label{}
  \int \, d\vr' \, \zbad (x') \Frt (\vr;\vr') = \cO (\eps^2)
\end{equation}
Indeed, according to \eqref{defFrt} and \eqref{scale}, the scaling change
$ \vr = \vrt / \kzd$ shows that 
\begin{align}
  \label{}
 \Frtt (\vrt;\vrt';\epsaap)=  &  
  \theta \left (\| \vrt-\vrt' \| - \kzd \sigma \right) 
  e^{-\epsaap \fdt (\vrt;\vrt')} -1 \nonumber \\
 &  + \epsaap \fdt (\vrt;\vrt')
   -\frac{1}{2} \epsaap^2 \left( \fdt (\vrt;\vrt') \right)^2 
\end{align}
with
  $\epsaap \equiv \kzd \beta e_\alpha e_\gamma$. Since $\zbad$ scales as 
$z_\alpha=\rho_{\alpha \, b} ^{\rm id}\propto  1/a^3$ (where $\rho_{\alpha \, b} ^{\rm id} $ 
is the bulk density of  an ideal gas with the same fugacities) and
\begin{equation}
  \label{}
\Ee \,\Frtt (\vrt;\vrt';\epsaap) = \cO (\epsaap^3), 
\end{equation}
$\int d\vr' \, \zbad(x') \Frt (\vr;\vr')$ is of order 
$\kzd^{-3} a^{-3} \epsaap^3 = \cO (\epsaap^2)$. (See definitions \eqref{defkdbis} 
and \eqref{defepsilon}.) We notice that the same result is obtained by using the scaling change 
$\vr = \beta e_\alpha e_\gamma \vu$ and the property that 
$\lim_{\eps \rightarrow 0} \Frt (\eps \vu,\eps \vu';\epsaap)$ is an integrable function of $\vu$ 
independent from $\eps$ so that \eqref{basicform} implies that 
$\int d\vr' \zbad (x') \Frt (\vr;\vr')$ is of order 
$(\beta e_\alpha e_{\alpha'} )^3 a^{-3} = \cO (\Gamma^3)  = \cO(\eps^2)$.
In fact, the lengths $b$ and $\sigma$ may also generate contributions of
order $\eps^2 \exp[\dw \beta e_\alpha^2/4b]$ or 
$\eps^2 \exp[\beta e_\alpha^2/\sigma]$, where $b$ and $\sigma$ 
are supposed to be finite.

A scaling analysis analogous to that performed in \cite{Cornu98II} shows
that more complicated diagrams contribute at least at the order $\eps^2$
in the exponential of the diagrammatic representation
\eqref{diagrhoresum}. 
Eventually, in order to calculate $\roa (x)$ up to
order $\eps$, in (\ref{diagrhoresum}) we only have to consider 
the contribution from the diagram with one bond $\Fcc$ up to order $\eps$.


\subsection{$\eps$-expansions of the contribution from the $\Fcc$ diagram}

In \eqref{diagrhoresum} the
contribution from $\Fcc$ up to order $\eps$ is derived from
\eqref{defFcc}, where $\fd$ is replaced by its leading value
$\fd^{(0)}$, and \eqref{crucialbis}. It is equal to the first order term in the 
$\eps$-expansion of  
\begin{multline}
  \label{contribFcc}
  - \frac{\beta e_\alpha  }{\kzd^2} \int_{\bt}^{+\infty} \, d\xt ' \,  
 \sum_\gamma z_\gamma e_\gamma
  \left[1+ \wzpg
\right]\\
  \left\{ \rule{0mm}{4mm} 1+\epsg \left[1-\Lb(\xt '; \bt, \dw ) \right] 
 +\Oexp(\eps^2)   
  \right\}\\
 \times \left\{ \phit^{(0)} (\xt,\xt ', \vq = {\bf 0};\bt) 
 +\phit^{(1)} (\xt , \xt ', \vq = {\bf 0};\bt, \dw)+ \Oexp(\eps^2)
  \right\}
\end{multline}
where $\bt = \kzd b$ and $\wza$ is defined in \eqref{defwzero}.
According to the neutrality constraint \eqref{zneut} and since the
function $\phit$ does not depend on the species $\gamma$, while the
integral of $f(\xt)$ times $\wza$ gives a contribution of order
larger than $\int d\xt f(\xt)$, the
contribution \eqref{contribFcc} involves only the leading term in the potential
$\fd$; it is reduced to 
\begin{multline}
\label{simpl}
- \frac{\beta e_\alpha }{\kdd} \int_{\bt}^{+\infty} \, d\xt ' \,
 \sum_\gamma z_\gamma e_\gamma
\left\{
\wzpg + \epsg\left[1-\Lb(\xt '; \bt, \dw)\right]
\right\}\\
\times \widetilde{\phi}^{(0)} (\xt, \xt', \vq = {\bf 0};\bt)
\end{multline}
where
$\widetilde{\phi}^{(0)} (\xt, \xt', \vq = {\bf 0};\bt)  = 2\pi\,[e^{-  | \xt - \xt' |} + e^{- (\xt+ \xt'-2\bt)}]$.

Equation \eqref{simpl} may be rewritten as 
\begin{multline}
  \label{titi}
 \sum_{\gamma} \int \, d\vr' \, \zbgd (x') \Fcc (\vr;\vr') = \\ 
 -  \frac{4 \pi \beta
  e_\alpha}{\kzd^2} \left( \sum_\gamma z_\gamma e_\gamma \eps_\gamma
  \left[ 1 - M_\gamma ( \kzd x; \eps_\gamma, \kzd b, \dw) \right] \right)
 + {\mathcal O}_{{\rm exp}} (\eps^2)
\end{multline}
where $M_\gamma =\bar{M} +[M_\gamma -\bar{M}]$ with 
\begin{equation}
  \label{exprMb}
  \bar{M} (\xt; \bt, \dw) = \frac{1}{2} \int_{\bt}^{+\infty} \, du' \,
 \left[ e^{-|\xt-u'|} + e^{-(\xt+u'-2\bt)} \right] \Lb (u';\bt,\dw) 
\end{equation}
and $M_\gamma - \bar{M}$ is the $\eps$-expansion at orders $\ln
\eps_\gamma$ and $(\eps_\gamma)^0$ of the integral 
\begin{equation}
  \label{ctu}
  - \frac{1}{2} \frac{1}{\eps_\gamma} \int_{\bt}^{+\infty} \, du' \,
  \left[ e^{-|\xt -u'|} + e^{-(\xt +u'-2\bt)} \right] w_0(u';
  \eps_\gamma, \dw)
\end{equation}


\subsection{Structure of the $\eps$-expansion for the density profile}

The structure of the $\eps$-expansion for the density profile can 
now be investigated. According to \eqref{diagrhoresum} and
the results of the previous section 
\begin{equation}
\label{expvaluerho}
\roa (x)=\zbad(x) \,\left[ 1+\sum_\gamma \int d\vr' \zbgd(x')
\Fcc(\vr,\vr';\alpha, \gamma)\right]\, 
\left[1+ \Oexp(\eps^2)\right]
\end{equation}
where $\zbad(x)$ is given by \eqref{crucialbis} and 
$\Fcc (\vr, \vr'; \alpha, \gamma)$ is replaced by 
$- \beta e_\alpha e_\gamma \kzd$  $\ftz  ( \kzd \vr, \kzd \vr')$. 
By using \eqref{titi} we get
\begin{align}
  \label{zexprho}
  \roa (x) = & \, \theta(x-b) \, z_\alpha  \, \exp \left( \dw \frac{\beta \ead}{4x} 
  e^{-2 \kd x} \right)
 \left\{ 1 \rule{0mm}{8mm} + \epsa \left[ 1 - \Lb \left( \kzd x; \kzd b,
  \dw \right) \right]  
  \right. \nonumber \\
 & \left. - \frac{ 4 \pi \beta e_\alpha}{\kzd^2} \sum_\gamma z_\gamma
  e_\gamma \eps_\gamma \left[ 1 - M_\gamma
  \left( \kzd x;\eps_\gamma, \kzd b,\dw \right) \right]  + \Oexp (\eps^2)
  \rule{0mm}{8mm}  \right\}
\end{align}

By definition of the bulk density \eqref{defrhob} and since $\Lb$ and $M_\gamma$
vanish when $x$ goes to infinity
\begin{equation}
  \label{zexprhob}
  \roab = z_\alpha \left\{ 1 + \epsa - \frac{\left( \sum_\gamma z_\gamma
  e_\gamma \eps_\gamma \right) 4 \pi \beta e_\alpha}{\kzd^2} +
  \mathcal{O}(\eps^2) \right\}
\end{equation}
The latter expression does coincide
with the classical limit of the particle density in a quantum plasma
where exchange effects are neglected (see Eq.(5.28) in Ref.\cite{ACP94}). 
Moreover, according to \eqref{defkdbis} and \eqref{defkzd}, $\kzd$ is
equal to $\kd$ up to a correction of order $\eps$
\begin{equation}
  \label{}
  \kzd = \kd \left[ 1 + \cO ( \eps ) \right]
\end{equation}
Using the property 
$\exp [ - ( 1 + \eps ) u ] = \exp [-u] - \eps u \exp[-u] + \cO \left(\eps^2 u \right) $,
comparison of \eqref{zexprho} with \eqref{zexprhob} leads to
\begin{align}
  \label{rhoexp}
  \roa(x) = & \roab \theta(x-b)   \exp \left( \dw \frac{\beta \ead}{4x} 
  e^{-2 \kd x} \right) \nonumber\\
 & \times \left\{ 1 \rule{0mm}{8mm} 
 - \frac{1}{2} \beta \kd \left[ \rule{0mm}{7mm}
  \ead \Lb 
 \left( \kd x; \kd b, \dw \right) \right.  \right. \nonumber \\
& \phantom{\times \left\{ 1 - \frac{1}{2} \beta \kd \left
  [ \rule{0mm}{7mm} \right. \right. } 
\left. - e_\alpha \frac{4
  \pi \beta }{\kd^2} \sum_\gamma \rogb e_\gamma^3 M_\gamma 
(\kd x;\eps_\gamma, \kd b,\dw) \right]  \nonumber \\
& \phantom{\times \left\{ 1 - \right. } \left. + \cO_{{\rm exp}} 
(\eps^2) \rule{0mm}{8mm} \right\}  
\end{align}
where $\epsg = (1/2) \beta \egd \kd$.

The term with $M_\gamma$ in \eqref{rhoexp} is related to the electrostatic 
potential drop $\Phi(x)$ created by the charge density profile. 
Indeed, when the relation between $\Phi (x)$ and the corresponding
electrostatic field is combined with Gauss theorem together with the
symmetries of the problem, and after an integration by parts, $\Phi (x)$
proves to read 
\begin{equation}
  \label{expVtot}
  \Phi (x) = - 4 \pi \int_x^{+\infty} dx' \, (x'-x) \soma \ea \roa (x') 
\end{equation}
(The 
value of $\Phi(x)$ when $x$ goes to $+\infty$ is chosen to be equal to zero).
When the structure \eqref{expvaluerho} of $\roa(x)$ is inserted into 
\eqref{expVtot}, according to \eqref{crucialbis} \eqref{basicformdeux} 
and \eqref{basicform}, the
contribution to $\Phi(x)$ at leading order $\eps$ only comes from the following
part in $\roa(x)$ (with $x>b$), 
\begin{equation}
  \label{}
  \zbad (x) + \za \, (-\beta \ea ) G(x) 
\end{equation}
where
\begin{equation}
  \label{defG}
  G(x) =  \int d\vrp \, \somg \eg \zbgd (x') \, \kzd \ftz (\kzd \vr, \kzd \vrp )
\end{equation}
Since $\ftz (\xt, \xt', \vq)$ obeys equation \eqref{eqphi1pwa}, 
\begin{equation}
  \label{}
  \kzd^2 \, G(x) = 4 \pi \somg \eg \zbgd (x) + \frac{d^2 G(x)}{d x^2} 
\end{equation}
when $x>b$.  Moreover $\kzd^2 =4\pi\beta\left(\soma  \za \ea^2\right)\times 
\left[1+\cO ( \eps )\right] $ so that
\begin{equation}
  \label{interm}
  \Phi(x) =\left[ \int_x^{+\infty} dx' (x'-x) \frac{d^2 G(x')}{d {x'}^2} \right]
  \left[1+\Oexp ( \eps )\right]
\end{equation}
According to the decay property of $\ftz$, $ G(x)$ tends to a constant exponentially 
fast when $x$ goes to $+\infty$. Thus, after integration by parts, 
\eqref{interm} leads to
\begin{equation}
  \label{genphi}
  \Phi(x) = \left[G(x) - \lim_{x \rightarrow + \infty} G(x)\right]
  \left[1+\Oexp ( \eps )\right]
\end{equation}
Since $-\beta e_\alpha G(x)$ coincides with the l.h.s. of \eqref{titi} at leading order in
$\eps$, the explicit value of $\Phi(x)$ 
is given by 
\begin{equation}
  \label{vtotcmlin}
  \Phi(x) = - \frac{2 \pi \beta}{\kd} \sum_\gamma \rog e_\gamma^3
  M_\gamma (\kd x; \eps_\gamma,  \kd b, \dw ) + \Oexp ( \eps^2 / \beta e )
\end{equation} 
The latter result ensures that the expression of the density profile derived in Paper 
I from a mean-field approach is exact at leading order in $\epsd$.


\appendix 

\section{Appendix A}
\label{AppendiceA}

In this appendix we show that, when $\uj (\xt)$ vanishes at large
 distances
 at least as
fast as $1/\xt$, $\Hjpe (\xt)$ defined in \eqref{defH} is  the only
 solution of the integral
equation \eqref{intrelHp}.
$\Hjpe (x; \rac, \bt)$ obeys the nonhomogeneous differential equation
\begin{equation}
  \label{eqH}
  \frac{d^2 \Hjp}{d\xt^2} - 2 \rac \frac{ d \Hjp}{d\xt} - \Uj (\xt) \Hjp =
  \Uj(\xt) 
\end{equation}
and satisfies the following boundary conditions 
\begin{equation}
  \label{bcondHb}
  \Hjpe (\bt; \rac, \bt) = 0
\end{equation}
and 
\begin{equation}
  \label{bcondHinf}
  \lim_{\xt \rightarrow + \infty} e^{-\xt \rac} \Hjpe (\xt; \rac, \bt ) = 0 
\end{equation}

By  using the extra  change of function
 $G_{{\rm j}} (\xt) = \exp \left[ -2 \xt \rac \right]$ \linebreak
  $\times d\Hjp/d\xt$ the problem is reduced to a 
first-order differential equation which relates 
$dG_i / d \xt $ to $H_j^+$. 
The formal integration of the latter equation leads to the relation 
\begin{equation}
  \label{relzero}
  \Hjp (\xt ; \rac , \bt ) = 
 a_0 + a_1 e^{ 2 \xt \rac} - \LUj \left[ 1 + \Hjp
  ; 2 \rac , \bt \right]
\end{equation}
where $\LUj$ is the linear operator defined in \eqref{defop}.

When $\uj$ tends to zero at least as fast as $U_0 / \xt$  when $\xt$
goes to infinity (with $U_0$ a constant), the boundary conditions
\eqref{bcondHb} and \eqref{bcondHinf} imply that the
integration constants $a_0$ and $a_1$ are in fact equal to zero in the
case of $\Hjpe$ . Indeed,
the hypothesis about $\uj$ means that there exists some $\xt_0$ such
that 
\begin{equation}
  \label{bornU}
  \forall \xt \geq \xt_0 \quad | \uj (\xt) | \leq \frac{U_0}{\xt}
\end{equation}
On the other hand, according to \eqref{bcondHinf}, for any given real
$M>0$ there exists some $\xt_1$ such that 
\begin{equation}
  \label{bornH}
  \forall \xt \geq \xt_1 (M) \quad e^{-\xt \rac} | 1 + \Hjpe (\xt) | < M
\end{equation}
Since $\exp ( \gamma v)$ is an increasing function of $v$, \eqref{bornU}
and \eqref{bornH} combined with properties of integrals lead to the
inequality 
\begin{multline}
  \label{bornun}
  \left| \LUj \left[ 1 + \Hjpe ; 2 \rac, \bt \right] (\xt) \right| 
  \leq {\mathcal
  L}_{| {\scriptscriptstyle U}_{\rm j} |} \left[ 1+ \Hjpe ; 2 \rac,
    \bt \right] (\xt_2) \\ + M U_0 e^{\xt \rac} \Lust \left[ 1; \rac, \xt_2
    \right] (\xt)
\end{multline}
with $\xt_2 = sup (\xt_0, \xt_1)$. In \eqref{bornun}, the left term in
the upper bound is independent from $\xt$ while the right term can be
calculated explicitly. An integration by parts leads to the result
\begin{equation}
  \label{fund}
  \Lust \left[ 1; \gamma, \bt \right] (\xt) = \frac{1}{\gamma} \left
  [ \ln (\gamma \xt) + B(\gamma \xt; \gamma \bt) \right]
\end{equation}
where $B(\gamma \xt; \gamma \bt)$ is a continuous function of $\xt$ which has
a finite limit when $\xt$ goes to $+\infty$ and which tends to $- \ln \bt$
when $\xt$ approaches the value $\bt$. Indeed,
\begin{equation}
  \label{defB}
  B(\gamma \xt; \gamma \bt) \equiv - \left[ e^{\gamma \xt} \Ei (- \gamma \xt) -
  e^{\gamma \bt} \Ei(-\gamma \bt) \right] - \ln (\gamma \bt)
\end{equation}
where $\Ei(-u)$ denotes the Exponential-Integral function defined for 
$u>0$ as $\Ei(-u) \equiv - \int_u^{+\infty} dt \, \exp(-t)/t$. As a
consequence of \eqref{bornun} and \eqref{fund}, in the large-$x$ limit
 $\exp(-2\xt \rac) \LUj \left[ 1+ \Hjp \right]$ vanishes as well as 
 $\Hjp$ \-  $\times \exp \left[ - 2 \xt \rac \right]$ 
(according to \-  \eqref{bcondHinf}) so that $a_1$ in \eqref{relzero}
proves to be zero. Then the definition \eqref{defop} together with the
choice \eqref{bcondHb} enforce that $a_0$ is also equal to zero.


\section{Appendix B}
\label{AppendiceB}

In the present appendix we study the structure of the $\eps$-expansions
of the functions $\HU$ defined by the series representation \eqref{series}.

\subsection{Bounds for the $\HUj$'s} 
\label{cdappenu}

According to  integration properties, the
linear operator $\LU$ defined in \eqref{defop} has the following properties,
\begin{equation}
  \label{vabsoluL}
  \left| \LU [f] \right| \leq \LabsU [ |f| ] 
\end{equation}
\begin{equation}
  \label{5.10modif}
  \left| U \right| < U' \, \Rightarrow \, \LabsU
  [|f|] \leq {\cal L}_{{\scriptscriptstyle U'}} [|f|]
\end{equation}
and
\begin{equation}
  \label{}
  0 \leq f \leq g \quad \Rightarrow \quad 0 \leq \LabsU [f] \leq \LabsU [g] 
\end{equation}
Therefore, by a recurrence argument, the series representation 
\eqref{series} of $\HU$ leads
to the following result
\begin{equation}
  \label{boundH}
  \mbox{ if} \,  \left| U \right| \leq U' \quad \mbox{ then} \quad \left|
  H_{{\scriptscriptstyle U}} \right| \leq H_{-{\scriptscriptstyle U'}}
\end{equation}

In the present appendix we will omit all irrelevant indices and
coefficients and we consider only the prototype functions
\begin{equation}
  \label{u1}
  U_1 (\xt;\eps) =  e^{ \dw \frac{\eps}{2 \xt}} - 1 
\end{equation}
and 
\begin{equation}
  \label{u2}
  U_2 (\xt; \eps ; \bt) = e^{ \eps F(\xt;\eps, \bt )} - 1
\end{equation}
where $F(\xt ; \eps, \bt)$ is a continuous function of $\xt$ in the interval 
$\bt \leq \xt < + \infty$,
which is bounded for all $\xt > \bt$ as follows, 
\begin{equation}
  \label{Bsbis}
  \left| F (\xt; \eps, \bt) \right| \leq M_F (\eps / \bt) g (\xt)
\end{equation}
where $g(\xt)$ is a continuous function integrable in the interval
$\left[ \bt, + \infty \right[$. Subsequently for all $\xt \geq \bt$
\begin{equation}
  \label{Bster}
  F(\xt; \eps, \bt) \leq M_F (\eps / \bt) M_g (\bt) \equiv M_2 (\eps/\bt, \bt)
\end{equation}
where $M_g (\bt)$ denotes the upper bound of $g$ in the range
 $\bt \leq \xt < + \infty$. 
 
Bounds upon the potentials $\uj$'s are 
derived by noticing that for all reals $v$ 
\begin{equation}
  \label{inegd}
  \left\{ \begin{array}{l}
 \left| e^{-|v|} - 1 \right| \leq |v| \\
  0  \leq  e^{|v|} -1 \leq |v| e^{|v|}
 \end{array} \right.
\end{equation}
(\eqref{inegd} is a direct consequence of the series representation of
the exponential.) \eqref{inegd} implies that 
\begin{equation}
  \label{boundUun}
  \left| U_1 (\xt; \eps) \right| < \Uumax \equiv \eps_1 \frac{1}{\xt}  
\end{equation}
where $\eps_1 = \eps M_1 (\eps/\bt, \dw)$
and
\begin{subequations}
\label{defMu}
\begin{alignat}{2}
M_1 (\eps / \bt, \dw) & = \frac{\dw}{2} \exp\left(\frac{\dw}{2}
  \frac{\eps}{\bt} \right) && \qquad \mbox{if} \quad \dw > 0 \\
& = \frac{|\dw|}{2} && \qquad \mbox{if} \quad \dw < 0 
\end{alignat}
\end{subequations}
Besides, if the sign of $F(\xt)$ depends on $\xt$, $F(\xt)$ may be written as
$F(\xt) = F_+ (\xt) - F_- (\xt)$ with $F_+ (\xt) = F(\xt)$ if $F(\xt) >0$ 
and $F_+(\xt) = 0 $ otherwise, while $F_- (\xt) = - F(\xt)$ if $F(\xt) < 0$ and 
$F_- (\xt) = 0 $ otherwise. With these definitions  
$|F(\xt)| = F_+(\xt) + F_-(\xt)$. Using the decomposition 
\begin{equation}
  \label{decomp}
  e^{\eps F} -1 = \left( e^{\eps F_+} -1 \right) e^{- \eps F_-} +
\left( e^{- \eps F_-} -1 \right)
\end{equation}
together with \eqref{Bsbis} and \eqref{inegd}, we get
\begin{equation}
  \label{boundUdeux}
  \left| U_2 (\xt; \eps, b ) \right| \leq \Udmax \equiv  \eps_2 
  g(\xt)  
\end{equation}
where $\eps_2 = \eps M_2 (\eps / \bt, \bt)$ and
\begin{equation}
  \label{defMd}
  M_2 (\eps / \bt, \bt) = M_F (\eps/ \bt) \exp \left[ \eps M_F (\eps /
  \bt) M_g (\bt) \right] 
\end{equation}
The bounds \eqref{boundUun} and \eqref{boundUdeux} about the 
$\bar{U}_{\rm  j}$'s  ensure that, according to \eqref{boundH}
\begin{equation}
  \label{bornHstar}
 \left| H_{U_1} \right| \leq H_{- \eps_1 /t} 
  \quad \mbox{and} \quad 
 \left| H_{U_2} \right|  \leq H_{-\eps_2 g}
\end{equation}
As shown in the following,
some information about the structure of the $\eps$-expansions of the
$\HUj$'s can be derived from  \eqref{bornHstar}.
In particular in both cases, the linearity of integrals ensures that the
$\nth$ term in the series \eqref{series} for $\HmUjmax$ is exactly of order
$\eps_j^n$, which is not the case when $U (\xt, \eps)$ is not a linear
function of $\eps$.


\subsection{$\eps$-expansions of bounds for the $\HUj$'s}
\label{B2}

First we consider $H_{-\eps g}$. A mere integration by parts allows one
to show that when $g(t)$ is positive, integrable at large $t$ 
and continuous for $\bt \leq t < + \infty$, 
\begin{equation}
  \label{Labsg}
  \Labsg [1] (\xt)  \leq \int_{\bt}^{+\infty} dt \,  g(t)  \equiv G(\bt) 
\end{equation}
Since $\Lmepsabsg [1] = -\eps \Labsg [1]$, a recurrence argument proves
that the $\nth$ term in the series representation
\eqref{series} of $H_{-\eps g}$ is a positive function of $x$ which is
lower than $\left[ \eps G(\bt) \right]^n$. Thus, 
$H_{-\eps g}(\xt)$ is lower than a geometric series which 
can be resummed with the result 
\begin{equation}
  \label{}
  H_{-\eps g}(\xt) \leq \frac{\eps G(\bt)}{1-\eps G(\bt)}
\end{equation}
Therefore $H_{-\eps g}(\xt ;\eps, \bt)$ can be written as an 
$\eps$-expansion whose coefficients are bounded functions of $\xt$,
\begin{equation}
  \label{cdB13}
  H_{-\eps g}(\xt) = \sum_{n=1}^{+\infty} \eps^n G_n (\xt;\bt)
\end{equation}
with $G_n (\xt; \bt) \leq G_n (\bt)$ for all $\xt \geq \bt$.

When $U(\xt) = \eps / \xt$ the starting equation \eqref{eqH} satisfied by
$\HU$, is exactly solvable and the formal series \eqref{series} must
coincide with the $\eps$-expansion of the explicit exact solution. More
precisely, $\exp(-x) \left[ 1 + \Hepsst(x) \right]$ is a solution of the
stationary radial Schroedinger equation for a quantum state with zero
angular kinetic momentum and a negative energy -1, in the Coulomb potential
$-\eps/r$ in dimensionless units, and with boundary conditions
\eqref{bcondHb} and \eqref{bcondHinf}. ( \eqref{bcondHb} is different
from the corresponding condition in a Hydrogen atom.) The equation reads 
\begin{equation}
  \label{eqcoul}
  \frac{d^2 h}{d\xt^2} + \left[ \frac{\eps}{\xt} -1 \right ] h = 0 
\end{equation}
The solutions of \eqref{eqcoul} are well-known : with the boundary
conditions \eqref{bcondHb} and \eqref{bcondHinf}
\begin{equation}
  \label{HW}
  1 + \Hepsst (\xt) = A(\eps, \bt) e^{\xt-\bt} \Wepssdusd (2\xt) 
\end{equation}
where $A(\eps, \bt) = \left[ \Wepssdusd (2\bt) \right]^{-1}$ and $W$ 
is the Whittaker function.  Up to a normalization factor 
\begin{equation}
  \label{whittak}
  \Wepssdusd (2 \xt) \varpropto 
 (2\xt)^{\eps/2}e^{-\xt} 
  \int_0^{+\infty} dt \, e^{-t}\, t^{-\eps/2}
   \left[ 1 + \frac{t} {2\xt}\right]^{\eps/2} 
\end{equation}

The structure of the $\eps$-expansion of $H_{\eps/t}$ can be derived
either from the $\eps$-expansion of the above integral representation of
$\Wepssdusd(2\xt)$ or from the series representation \eqref{series}. 
  (\eqref{series} is
also an $\eps$-expansion of $H_{\eps/t}$, since the $n^{{\rm th}}$ 
term is exactly
of order $\eps^n$ because of the linearity property of integration.)
We choose to exhibit the
second derivation in order to illustrate how the series representation
provides information.  
A recurrence scheme using an integration by parts
and the fact that 
${\cal L}_{1/t}[f(\gamma t);\gamma,\bt](\xt)=(1/\gamma){\cal L}_{1/t}[f(t);1,\gamma\bt](\gamma\xt)$
allows one to show a generalization of \eqref{fund} (with $\gamma = 2$),
\begin{equation}
  \label{}
  \Lust \left[ \left( \ln 2 t \right)^p ; 2, \bt \right] (\xt) =
  \frac{1}{2(p+1)} \left( \ln 2 \xt \right)^{p+1} 
  + \sum_{n=1}^{p} A_{p, n}
  \left( \ln 2 \xt \right)^n + B_p (\xt,\bt)
\end{equation}
where $B_p(\xt,\bt)$ is a bounded function of $\xt$ when
 $\xt>\bt$. Since  the $n^{{\rm th}}$  term in the series \eqref{series} for
$U=\eps/t$ is exactly proportional to $\eps^n$, a recurrence scheme
readily shows that the structure of the $\eps$-expansion of $\Hepsst (\xt)$
reads
\begin{equation}
  \label{cdapBdn}
  \Hepsst (\xt) = \sum_{n=1}^{+\infty} \eps^n \left[ \frac{1}{n!} \left
  ( \frac{\ln 2 \xt }{2} \right)^n  + b_{n-1} \left( \ln 2 \xt \right)^{n-1}
    + \ldots +
  b_1 \ln (2 \xt) + B_n^*(\xt,\bt) \right]
\end{equation}
In \eqref{cdapBdn} $B_n^* (\xt,\bt)$ is a function which is bounded 
for $\xt>\bt$.
The coefficient at every order $\eps^n$
vanishes when $\xt=\bt$ (because $\Lust [f] (\xt=\bt) = 0 $). It
diverges as $(1/n!) \left(\ln (2\xt) / 2 \right)^n$ at
large $\xt$ and the summation over the leading asymptotic behaviours
at all orders in $\eps$ can be performed explicitly with the result
\begin{equation}
  \label{Has}
  \Hepsst ( \xt ) \underset{\xt \rightarrow + \infty}{\sim} e^
  { \frac{\eps}{2} \ln (2\xt)} = (2\xt)^{\eps / 2}
\end{equation}
The asymptotic behaviour \eqref{Has} indeed coincides with the leading
large-distance behaviour of the explicit solution \eqref{HW} 
derived by using the integral representation \eqref{whittak}, 
\begin{equation}
  \label{Was}
 \Wepssdusd (2\xt) \underset{ \xt \rightarrow + \infty}{\sim} \left( 2 \xt
 \right)^{\eps/2} e^{-\xt} \left[ 1 - \frac{\eps(2+\eps)}{8\xt} +
 \mathcal{O} \left(\frac{1}{\xt^2}\right) \right] 
\end{equation}


\subsection{Exact solution at first order in    $\eps$}

First, the fact that the upper bounds $\Ujmax (\xt;\eps_{{\rm j}},\bt)$'s 
upon the $U_{{\rm j}}$'s are of the form 
\begin{equation}
  \label{Umax}
  \Ujmax (\xt;\eps_{{\rm
    j}},\bt) = \eps_{{\rm j}} V_{{\rm j}}^{{\rm max}} (\xt)
\end{equation}
(where $\eps_1$ and $\eps_2$ are defined in \eqref{boundUun} and
\eqref{boundUdeux}) allows one to show that  
 \begin{equation}
  \label{Hexp}
  \HUj = - \LUj [1] + \sum_{n=2}^{+\infty} \eps^n a_{{\rm j},n}
  (\xt;\eps, \bt) 
\end{equation}  
with for all $\xt \geq \bt$, 
\begin{equation}
  \label{B23bis}
  \left| a_{{\rm j},n} (\xt;\eps, \bt)  \right| \leq \left[ M_{{\rm j}}
  (\eps / \bt, \bt) \right]^n A_{{\rm j}, n} (\xt, \bt^{j-1})
\end{equation}
where the $M_{{\rm j}}$'s are defined in \eqref{defMu} and
\eqref{defMd}.
In \eqref{B23bis} 
 $A_{1,n} (x;\eps,b)$ does not diverge faster than $(\ln \xt)^n$ when
$\xt$ goes to infinity, while $A_{2,n}(\xt;\bt)$ is   a bounded function
of $\xt$ for all $\xt \geq \bt$. 
The demonstration is the following. According to the series
representation \eqref{series} for $\HUj$, the inequalities
    \eqref{vabsoluL} and \eqref{5.10modif} together with the
 proportionality relation \eqref{Umax} lead to 
    \begin{equation}
      \label{Bun}
      \left| \HUj + \LUj [1] \right| \leq \sum_{n=2}^{+\infty} \eps_j^n
      \LUjbmax \left[ \LUjbmax \cdots \left[ \LUjbmax [1] \right] \cdots
      \right] 
    \end{equation}
where the coefficient of $\eps_j^n$ contains $n$ operators $\LUjbmax$.
\eqref{Bun} implies that the sum in \eqref{Hexp} indeed starts at the
order $\eps^2$. Moreover, according to \eqref{series},  
the right term in \eqref{Bun} may be written as 
\begin{equation}
  \label{}
  \epsj^2 \LUjbmax \left[ \LUjbmax \left[ \HmUjmax + \LmUjmax [1] \right]
  \right]
\end{equation}
The explicit structures of the $\eps_{{\rm j}}$-expansions \eqref{cdB13} and
 \eqref{cdapBdn} for the \linebreak $\HmUjmax$'s together with the results
 \eqref{fund} and \eqref{Labsg} for $\LmUjmax$ imply that 
\begin{equation}
   \label{}
   \left| \HUj + \LUj [1] \right| \leq \sum_{n=2}^{+\infty} \eps_{{\rm
   j}}^n A_{{\rm j} ,\, n} (\xt; \bt^{j-1})
 \end{equation}
where the $A_{{\rm j} , \, n}$ have the properties given after
\eqref{B23bis}.

Second, we show that the $\eps$-expansion of
$\LUj [1] - \LUjlin [1]$ starts at order $\eps^2$,
where the linearized potential
$\Ujl$ is the first-order term in the expansion of the
exponential involved in the definitions \eqref{u1} and \eqref{u2} of the
$\Uj$'s. More precisely
\begin{equation}
  \label{Lexp}
  \LUj [1] = \LUjlin [1] + \eps^2 \RLj (\xt;\eps,\bt) 
\end{equation}
where for all $\xt \geq \bt$, $\RLj$ is a bounded function of $\xt$ and
the dependence of its bound $M_{L_{{\rm j}}} (\eps/ \bt, \bt)$ upon
$\eps$ is entirely contained in $M_1 (\eps/b)$ if $j=1$ or in 
$[ M_F (\eps/ \bt)]^2 \-  \exp [\eps M_F (\eps / \bt) \-  M_g (\bt)]$ 
if $j=2$.
\eqref{Lexp} is derived as follows. First we notice that 
\begin{equation}
  \label{borndifL}
  \LUj [1] - \LUjlin [1] = \LdifU [1]
\end{equation}
Moreover for all reals $v$
\begin{equation}
  \label{}
  \left\{
  \begin{array}{l}
  \left| e^{-|v|} - 1 + |v| \right| \leq v^2 \\
  e^{|v|} -1 -|v| \leq v^2 e^{|v|}
  \end{array}
  \right.
\end{equation}
so that 
\begin{equation}
  \label{}
 \left|  U_1 (\xt;\eps) - U_1^{ \, {\rm lin}}(\xt;\eps) \right| \leq 
 \frac{|\dw|}{2} M_1 (\eps/ \bt) \eps^2  \frac{1}{x^2} 
\end{equation}
A decomposition similar to \eqref{decomp} reads
\begin{multline}
  \label{}
  e^{\eps F} - 1 - \eps F = 
 \left( e^{\eps F_{+}} -1 -\eps F_{+} \right)
  e^{-\eps F_{-}} + \eps F_{+} \left( e^{-\eps F_{-}} - 1 \right) \\ + 
 \left( e^{-\eps F_{-}} - 1 +\eps F_{-} \right) 
\end{multline}
and we get
\begin{equation}
  \label{}
  \left| U_2 (\xt;\eps) - U_2^{ \, {\rm lin}}(\xt;\eps) 
 \right| \leq M_F (\eps/\bt) M_2 (\eps/\bt, \bt)
\eps^2 g^2 (\xt) 
\end{equation}
According to \eqref{5.10modif}, the latter bounds together with
\eqref{borndifL} imply that 
\begin{equation}
  \label{}
  \left| \LUj [1] - \LUjlin [1] \right| \leq \eps^2 \Lgj [1]
\end{equation}
where $g_1 (\xt) = (|\dw|/2) M_1 (\eps/\bt)  1/x^2$ and
$g_2 (\xt) = M_F (\eps/\bt) M_2 (\eps/ \bt, \bt) g(\xt)^2$. 
Since $1/\xt^2$ as well as $g(\xt)^2$ are continuous for 
$\xt \geq \bt$ and integrable when $\xt$ goes to $\infty$, 
 the result \eqref{Labsg} can be applied to the $g_j$'s and
leads to \eqref{Lexp}. 

Subsequently, according to \eqref{Hexp} and \eqref{Lexp}
\begin{equation}
  \label{final}
  \HUj (\xt) = - \LUjlin [1] (\xt) + \eps^2 \RUj (\xt; \eps, \bt )
\end{equation}
where if $\eps/\bt$ and $\bt$ are kept fixed, 
$\lim_{\eps \rightarrow 0} \RUj (\xt, \eps, \bt) < + \infty$ and 
$\RUj (\xt, \eps, \bt)$ does not diverge faster than $\xt^{p(\eps)}$ with 
$\lim_{\eps \rightarrow 0} p(\eps)$ \- $  \! \! = 0 $ when $\xt$ goes to $+\infty$. 
We recall that the result \eqref{final} holds for a potential $\Uj (\xt;\eps)$ which
may be not bounded for all $\xt \geq 0$. The only hypotheses about 
$\Uj (\xt; \eps)$ are that $\Uj (\xt; \eps)$ tends to zero at least as fast as
$\eps/\xt$ when $\xt$ becomes larger than 1.
The property \eqref{final} can be applied to $\Ujlin$ and we get 
\begin{equation}
  \label{}
  \HUj (\xt) - \HUjlin (\xt) = \eps^2 \left[ \RUj (\xt;\eps, \bt) - 
 \RUjlin (\xt; \eps, \bt) \right]
\end{equation}
The solutions of the homogeneous equations \eqref{eqH} for $\Uj$ and
$\Ujlin$ respectively coincide only at the first order in their
$\eps$-expansions.

\bibliographystyle{plain}
\bibliography{biblio}


\begin{figure}
\centering\epsfig{figure=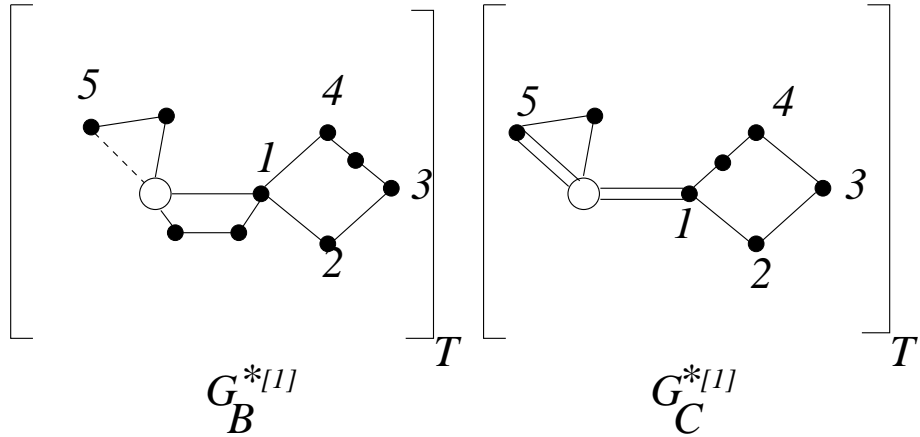, width=12cm}
\caption{Step-2 for resummations. $G_B^{* \, [1]}$ and $G_C^{* \, [1]}$
  are two $G^{* \, [1]}$ diagrams which correspond to the same prototype
  diagram $\cpp$ as the diagram $G_A^{* \, [1]}$ in Fig. \ref{f2}. The
  brackets with an index $T$ indicate that the diagrams $G^{* \, [1]}$
  must be calculated with the weights and the subtraction (rule $\rr$)
  displayed in Fig. \ref{f2}
} \label{f3}
\end{figure}

\begin{figure}
\centering\epsfig{figure=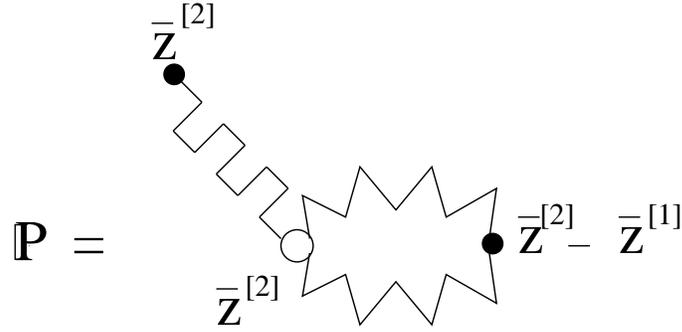, width=9cm}
\caption{Diagram $\cpp$ corresponding to $G_A^{* \, [1]}$, 
$G_B^{* \,    [1]}$ and $G_C^{* \, [1]}$ after step-2 resummation. 
The crenelated line is a bond $\Frt$, while the double wavyline 
corresponds to a bond  $(1/2) \left[ \Fcc \right]^2$. 
The rule $\rr_2$ is illustrated in the present example. 
} \label{f4}
\end{figure}

\begin{figure}
\centering\epsfig{figure=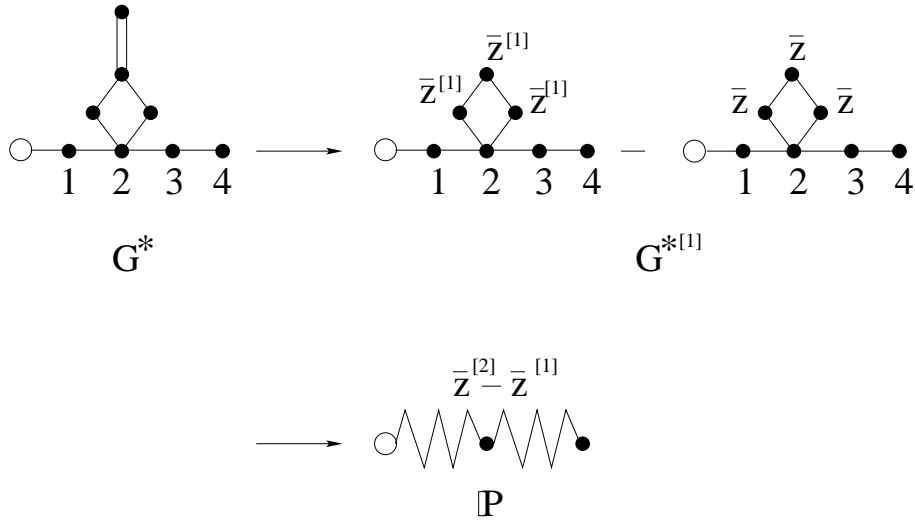, width=12cm}
\caption{Example for rule $\rr_1$. Diagram $G^*$ contributes to the
  value of diagram $G^{* \, [1]}$ after step-1 resummations. The value of
  $G^{* \, [1]}$ is determined by rule $\rr$. After step-2 resummations
$G^{* \, [1]}$ contributes to diagram $\cpp$,  whose value is given by rule
  $\rr_1$. A single wavy line denotes a bond $\Fcc$.
} \label{f5}
\end{figure}

\end{document}